# Early Release Science of the Exoplanet WASP-39b with JWST NIRSpec G395H

Lili Alderson[1*], Hannah R. Wakeford[1*], Munazza K. Alam[2], Natasha E. Batalha[3], Joshua D. Lothringer[4], Jea Adams Redai[5], Saugata Barat[6], Jonathan Brande[7], Mario Damiano[8], Tansu Daylan[9,10], Néstor Espinoza[11,12], Laura Flagg[13], Jayesh M. Goyal[14], David Grant[1], Renyu Hu[8,15], Julie Inglis[15], Elspeth K. H. Lee[16], Thomas Mikal-Evans[17], Lakeisha Ramos-Rosado[12], Pierre-Alexis Roy[18], Nicole L. Wallack[2,15], Natalie M. Batalha[19], Jacob L. Bean[20], Björn Benneke[18], Zachory K. Berta-Thompson[21], Aarynn L. Carter[19], Quentin Changeat[22,23], Knicole D. Colón[24], Ian J.M. Crossfield[7], Jean-Michel Désert[6], Daniel Foreman-Mackey[25], Neale P. Gibson[26], Laura Kreidberg[17], Michael R. Line[27], Mercedes López-Morales[5], Karan Molaverdikhani[28,29], Sarah E. Moran[30], Giuseppe Morello[31,32,33], Julianne I. Moses[34], Sagnick Mukherjee[19], Everett Schlawin[35], David K. Sing[12,36], Kevin B. Stevenson[37], Jake Taylor[38,18], Keshav Aggarwal[39], Eva-Maria Ahrer[40,41], Natalie H. Allen[12,42], Joanna K. Barstow[43], Taylor J. Bell[44], Jasmina Blecic[45,46], Sarah L. Casewell[47], Katy L. Chubb[48], Nicolas Crouzet[49], Patricio E. Cubillos[50,51], Leen Decin[52], Adina D. Feinstein[20,42], Joanthan J. Fortney[19], Joseph Harrington[53], Kevin Heng[41,54], Nicolas Iro[55], Eliza M.-R. Kempton[56], James Kirk[5,57,58], Heather A. Knutson[15], Jessica Krick[59], Jérémy Leconte[60], Monika Lendl[61], Ryan J. MacDonald[62,13,63], Luigi Mancini[64,65,17], Megan Mansfield[35,63], Erin M. May[37], Nathan J. Mayne[66], Yamila Miguel[49,67], Nikolay K. Nikolov[11], Kazumasa Ohno[19], Enric Palle[31], Vivien Parmentier[38,68], Dominique J. M. Petit dit de la Roche[61], Caroline Piaulet[18], Diana Powell[5,63], Benjamin V. Rackham[69,70,71], Seth Redfield[72], Laura K. Rogers[73], Zafar Rustamkulov[12], Xianyu Tan[38], P. Tremblin[74], Shang-Min Tsai[38], Jake D. Turner[13,63], Miguel de Val-Borro[75], Olivia Venot[76], Luis Welbanks[27,63], Peter J. Wheatley[40,41], Xi Zhang[77]

*Corresponding authors' emails: lili.alderson@bristol.ac.uk, hannah.wakeford@bristol.ac.uk
All author affiliations are listed at the end of the paper

**Measuring the abundances of carbon and oxygen in exoplanet atmospheres is considered a crucial avenue for unlocking the formation and evolution of exoplanetary systems[1,2]. Access to an exoplanet's chemical inventory requires high precision observations, often inferred from individual molecular detections with low-resolution space-based[3-5] and high-resolution ground-based[6-8] facilities. Here we report the medium-resolution (R≈600) transmission spectrum of an exoplanet atmosphere between 3–5 µm covering multiple absorption features for the Saturn-mass exoplanet WASP-39b[9], obtained with JWST NIRSpec G395H. Our observations achieve 1.46× photon precision, providing an average transit depth uncertainty of 221 ppm per spectroscopic bin, and present minimal impacts from systematic effects. We detect significant absorption from $CO_2$ (28.5σ) and $H_2O$ (21.5σ), and identify $SO_2$ as the source of absorption at 4.1 µm (4.8σ). Best-fit atmospheric models range between 3× and 10× solar metallicity, with sub-solar to solar C/O ratios. These results, including the detection of $SO_2$, underscore the importance of characterising the chemistry in**

**exoplanet atmospheres, and showcase NIRSpec G395H as an excellent mode for time series observations over this critical wavelength range[10].**

We obtained a single transit observation of WASP-39b using the Near Infrared Spectrograph (NIRSpec)[11,12] G395H grating on 30-31 July 2022 between 21:45 - 06:21 UTC using the Bright Object Time Series mode (BOTS). WASP-39b is a hot ($T_{eq}$ = 1120 K) low density giant planet with an extended atmosphere. Previous spectroscopic observations have revealed prominent atmospheric absorption by Na, K, and $H_2O$[3,4,13-15], with tentative evidence of $CO_2$ from infrared photometry[4]. Atmospheric models fitted to the spectrum have inferred metallicities (amount of heavy elements relative to the host star) from 0.003 – 300x solar[3,15-20], which makes it difficult to ascertain the planet's formation pathway[21,22]. The host, WASP-39, is a G8 type star which displays little photometric variability[23], and has nearly solar elemental abundance patterns[24]. The quiet host and extended planetary atmosphere make WASP-39b an ideal exoplanet for transmission spectroscopy[25]. The transmission spectrum of WASP-39b was observed as part of the JWST Transiting Exoplanet Community Director's Discretionary Early Release Science (JTEC ERS) program[26,27] (ERS-1366 PIs Natalie M. Batalha, Jacob L. Bean and Kevin B. Stevenson) which uses four instrument configurations to test their capabilities and provide lessons learned for the community.

The NIRSpec G395H data was recorded with the 1.6" × 1.6" fixed slit aperture using the SUB2048 subarray and NRSRAPID readout pattern, with spectra dispersed across both the NRS1 and NRS2 detectors. Over the ~8-hour duration of the observation, a total of 465 integrations were taken, centred around the 2.8-hour transit. We obtained 70 groups per integration, resulting in an effective integration time of 63.14 seconds. During the observation, the telescope experienced a "tilt event", a spontaneous and abrupt change in the position of one or more mirror segments, causing changes in the point spread function (PSF) and hence jumps in flux[28]. The tilt event occurred mid-transit, impacting ~3 integrations, and resulted in a noticeable step in the flux time-series, the size of which is wavelength-dependent (Figure 1 and Methods). The tilt event also affects the PSF, with the full-width-at-half-maximum (FWHM) of the spectral trace displaying a step-function-like shape (see Extended Data Figures 2 and 3).

We produced multiple reductions of the observations using independent analysis pipelines (see Methods). For each reduction, we created broadband and spectroscopic light curves from 2.725–3.716 μm for NRS1 and 3.829–5.172 μm for NRS2 using 10-pixel wide bins (≈0.007 μm, median resolution R≈600), excluding the detector gap between 3.717–3.823 μm. The light curves show a settling ramp during the first 10 integrations (≈631.4 seconds), with a linear slope across the entire observation for NRS1. We otherwise see no substantial systematic trends and achieve improvements in precision from raw uncorrected to fitted broadband light curves of 1.63× to 1.03× photon noise for NRS1 and 1.95× to 1.31× for NRS2. The flux jump caused by the mirror tilt event could be corrected by detrending against the spectral trace x and y positional shifts, normalising the light curves, or fitting the light curves with a step function (see Methods). We produced multiple fits from each set of light curves, resulting in a total of 11 independently measured transmission spectra. Figure 1 demonstrates that our spectroscopic light curves achieve precisions close to photon noise, with a median precision of 1.46× photon noise across the full wavelength range (see Extended Data Figure 4).

We show transmission spectra from multiple combinations of independent reductions and light curve fitting routines in Figure 2, along with the weighted average of all 11 transmission spectra with the unweighted mean uncertainty produced by our analyses (see Methods). We find that using different combinations of reduction and fitting methods result in consistent transmission spectra (see Methods and Extended Data Figure 5). While we see some artefacts at the edges of the detectors (see Figure 3, bottom panel) which may be caused by uncharacterised systematics, these only impact a small number of wavelength bins. Our resulting averaged NIRSpec G395H spectrum shows increased absorption towards bluer wavelengths short of 3.7 μm and a prominent absorption feature between 4.2–4.5 μm, along with a smaller amplitude absorption feature at 4.1 μm and a narrow feature around 4.56 μm.

We compared the weighted average G395H transmission spectrum to three grids of 1D radiative-convective thermochemical equilibrium (RCTE) atmosphere models of WASP-39b (see Methods, Extended Data Table 2), containing a total of 10,308 model spectra. The best-fit models from each grid provide a reduced chi-square per data point ($\chi^2/N$) of 1.08–1.20 for our 344 data point transmission spectrum (Figure 3). The increased absorption at blue wavelengths across NRS1 is consistent with absorption from $H_2O$ (at 21.5σ, see Methods), while the large bump in absorption between 4.2–4.5 μm[29] can be attributed to $CO_2$ (28.5σ). $H_2O$ and $CO_2$ are expected atmospheric constituents for near-solar atmospheric metallicities, with the $CO_2$ abundance increasing non-linearly with higher metallicity[30]. The spectral feature at 4.56 μm (3.3σ) is currently unidentified, but does not correlate with any obvious detector artefacts and is reproduced by multiple independent analyses. The absorption feature at 4.1 μm is also not seen in the RCTE model grids. After an exhaustive search for possible opacity sources[31], described in the corresponding NIRSpec PRISM analysis[32], we interpret this feature as $SO_2$ (4.8σ), as it is the best candidate at this wavelength.

While $SO_2$ would have volume mixing ratios (VMR) of less than $10^{-10}$ throughout most of the observable atmosphere in thermochemical equilibrium, coupled photochemistry of $H_2S$ and $H_2O$ can produce $SO_2$ on giant exoplanets, with the resulting $SO_2$ mixing ratio expected to increase with increasing atmospheric metallicity[33–35]. We find that approximately $\sim 10^{-6}$ VMR of $SO_2$ is required to fit the spectral feature at 4.1 μm in WASP-39b's transmission spectrum, consistent with lower-resolution NIRSpec PRISM observations of this planet[32] and previous photochemical modelling of super-solar metallicity giant exoplanets[35,36]. Figure 4 shows a breakdown of the contributing opacity sources for the lowest $\chi^2/N$ best-fit model (PICASO 3.0) with VMR=$10^{-5.6}$ injected $SO_2$. The inclusion of $SO_2$ in the models results in an improved $\chi^2/N$ and is detected at 4.8σ (see Methods), confirming its presence in the atmosphere of WASP-39b.

We additionally look for evidence of $CH_4$, CO, $H_2S$, and OCS (carbonyl sulfide) because their near-solar chemical equilibrium abundances could result in a contribution to the spectrum.

We see no evidence of $CH_4$ in our spectrum between 3–3.6 μm[23], which is indicative of C/O < 1[37] and/or photochemical destruction[35,38]. With regards to CO, $H_2S$ and OCS, we were unable to conclusively confirm their presence with these data. In particular, CO, $H_2O$, OCS, and our modeled cloud deck all have overlapping opacity, which creates a pseudo-continuum from 4.6–5.1 μm (see Figures 3 and 4). Therefore, we were unable to unambiguously identify the individual contributions from CO and other molecules over this wavelength region at the resolution presented in this work.

Our models reveal an atmosphere enriched in heavy elements, with best-fit parameters ranging from 3–10× solar metallicity, given individual model grid's spacing (see Methods). The spectra also indicate C/O ratios ranging from sub-solar to solar depending on the grid used, informed by the relative strength of absorption from carbon-bearing molecules to oxygen-bearing molecules. The interpretation of the relatively high resolution and precision of the G395H spectrum appears to be sensitive to the treatment of aerosols in the model, with one grid preferring 3× solar metallicity when using a wavelength-dependent cloud opacity and physically-motivated vertical cloud distribution[39], but 10× solar metallicity when assuming a grey cloud. In general, forward model grids fit the major features of the data, but do not place statistically significant constraints on many of the atmospheric parameters (see Methods). Future interpretation of the JTEC ERS WASP-39b data with Bayesian retrieval analyses will provide robust confidence intervals for these planetary properties and explore the degree to which these data are sensitive to modelling assumptions (e.g., chemical equilibrium versus disequilibrium) and parameter degeneracies (e.g., clouds versus atmospheric metallicity).

We are able to strongly rule out an absolute C/O ≥ 1 scenario ($\chi^2/N \geq 3.97$), which has previously been hypothesised for gas-dominated accretion at wide orbital radii beyond the $CO_2$ ice line where the gas may be carbon-rich[40]. Our C/O upper limit, therefore, suggests that WASP-39b may have either formed at smaller orbital radii with gas-dominated accretion or that the accretion of solids enriched WASP-39b's atmosphere with oxygen-bearing species[2]. The level of metal enrichment (3–10× solar) is consistent with similar measurements of Jupiter and Saturn[41,42], potentially suggesting core-accretion formation scenarios[43], and is consistent with upper limits from interior structure modelling[44]. These NIRSpec G395H transmission spectroscopy observations demonstrate the promise of robustly characterising the atmospheric properties of exoplanets with JWST unburdened by significant instrumental systematics, unravelling the nature and origins of exoplanetary systems.

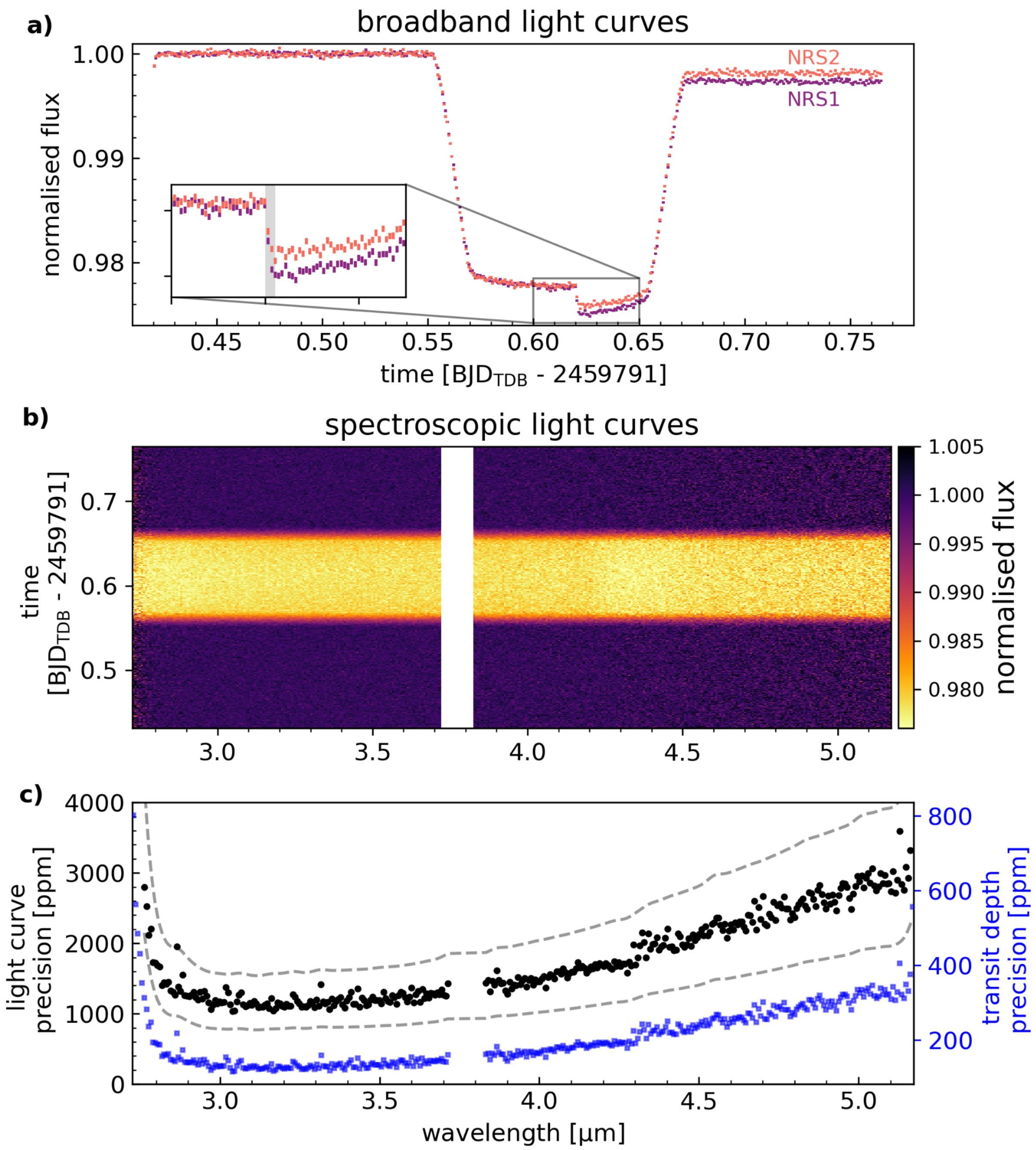

**Figure 1: Light-curve precisions achieved for WASP-39b with NIRSpec G395H.** a) raw, uncorrected broadband light curves from the NRS1 (purple) and NRS2 (red) detectors, demonstrating the lack of dominant systematic trends in the light curves. The inset shows the drop in flux (grey-shaded region) caused by a mirror tilt event, resulting in a distinct change in flux between NRS1 and NRS2 after the tilt event (see Extended Data Figures 2 and 3). b) pixel intensity map of the spectroscopic light curves after correction for the tilt event and additional instrument systematics. c) light curve precision obtained per spectroscopic bin (black) compared to 1x and 2x photon noise expectations (grey dashed lines) and the measured precision on the transit depth (blue). The gap between the two detectors (3.72–3.82 μm) is highlighted in the middle and bottom plots. All data shown are from fitting pipeline 1 (see Methods).

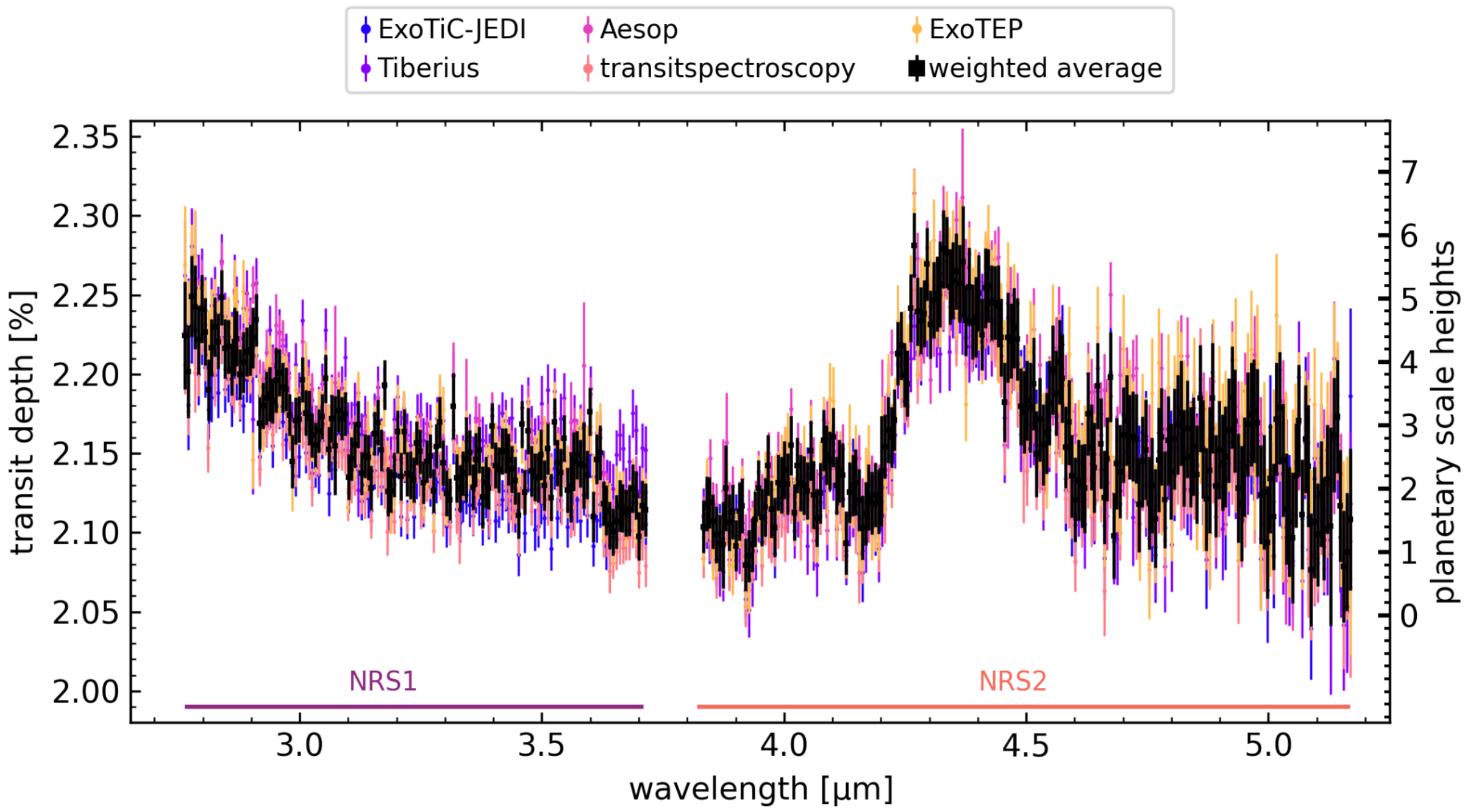

**Figure 2: WASP-39b transmission spectra measured at 10-pixel resolution (≈7 nm wide bins, R≈600) using multiple fitting pipelines.** We show the resultant spectra from five out of eleven independent fitting pipelines, which used distinct analysis methods to demonstrate the robust structure of the spectrum (see Methods for details on each fitting pipeline and comparative statistics). The black points show the weighted average transmission spectrum computed from the transit depth values in each bin weighted by $1/\sigma^2$, where σ is the uncertainty on the data point from each of the 11 fitting pipelines. The error bars were computed from the unweighted mean uncertainty in each bin (see Extended Data Figure 5). All spectra show consistent broadband absorption short of 3.7 μm, around 4.1μm, and from 4.2 to 4.5 μm.

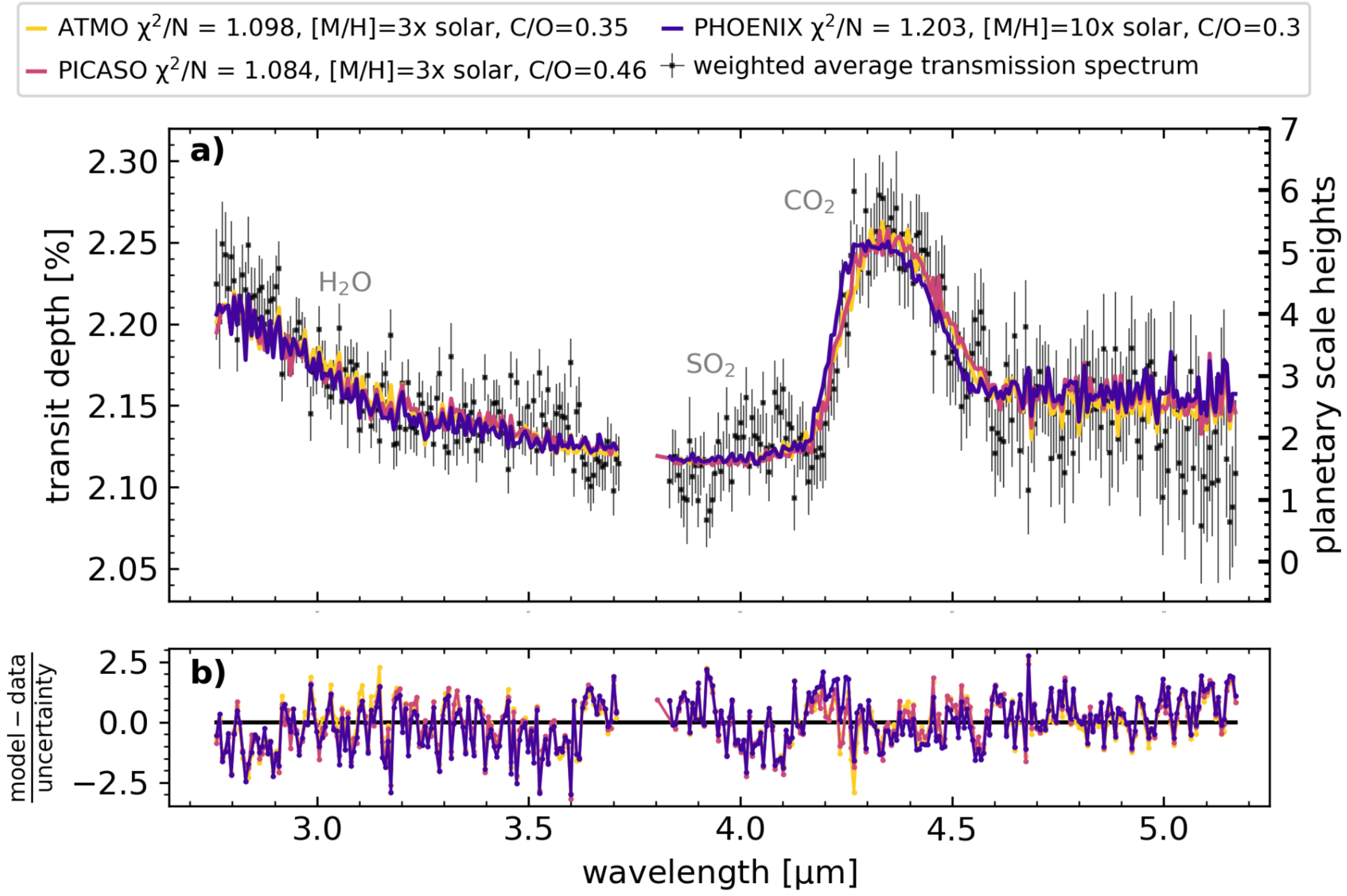

**Figure 3: Spectra from three independent 1D Radiative Convective Thermochemical Equilibrium (RCTE) models (a) and their residuals (b), fit to the weighted average WASP-39b G395H transmission spectrum.** The models are dominated by absorption from $H_2O$ and $CO_2$ with a grey cloud top pressure corresponding to ≈1 mbar. The models find that the data are best explained by 3–10× solar metallicity and sub-solar to solar C/O (C/O = 0.3-0.46). The additional absorption due to $SO_2$, seen in the spectrum around 4.1 μm, is not included in the RCTE model grids, and causes a significant impact on the $\chi^2$/N (see Figure 4).

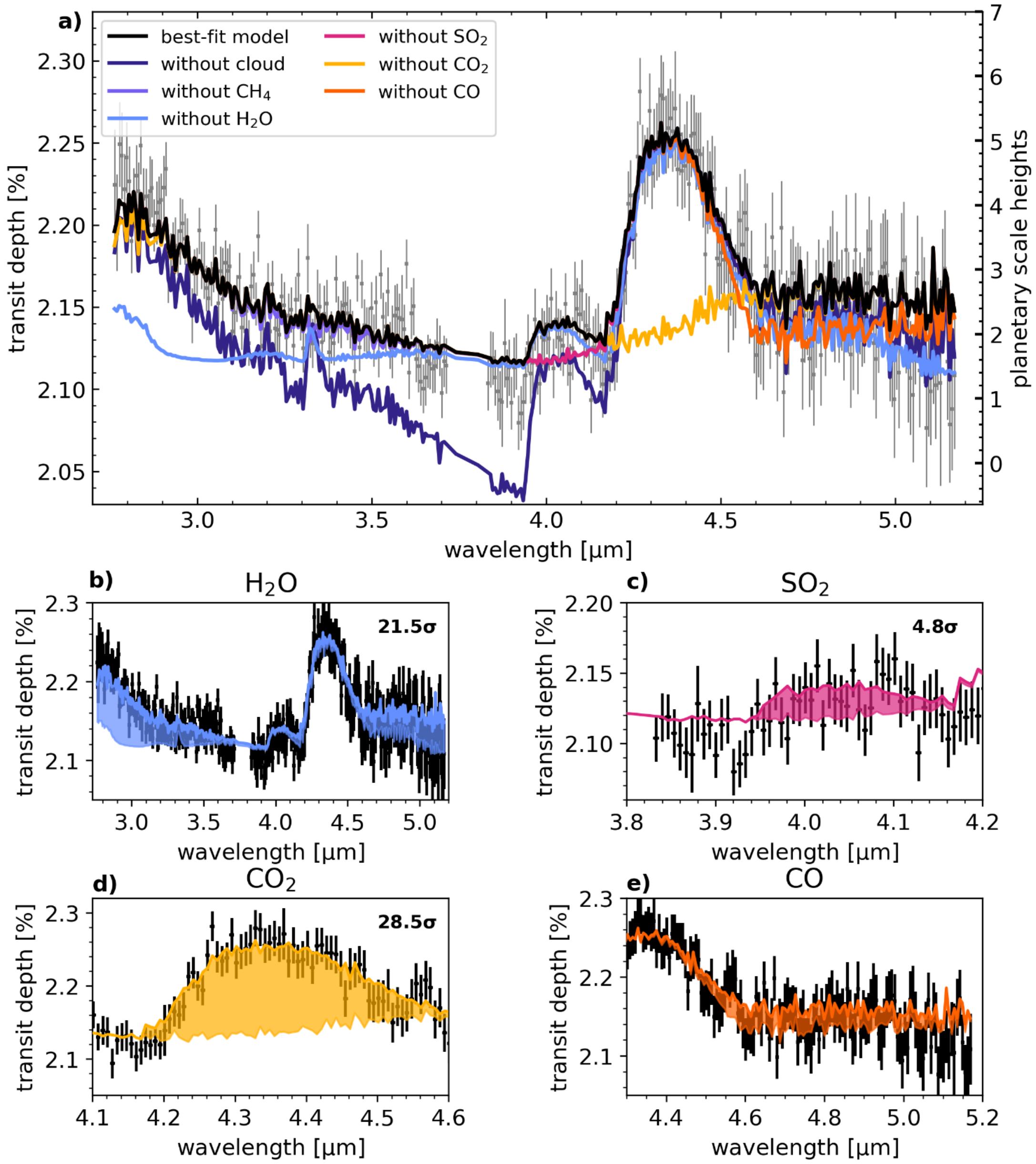

**Figure 4: Contribution of opacity sources to the best-fitting model with injected $SO_2$.** a) show the lowest $\chi^2/N$ best-fitting model (PICASO in Figure 3) with an injected abundance of $10^{-5.6}$ (VMR) $SO_2$. We also show this model with a selection of the anticipated absorbing species and the cloud opacity removed to indicate their contributions to the model. The inclusion of $SO_2$ in the model decreases the $\chi^2/N$ from 1.08 (shown in Figure 3) to 1.02, resulting in a 4.8σ detection (see Extended Data Table 3). b) to e) show the effect of removing the corresponding molecular opacity from the spectrum (shaded region). Our best-fit model is additionally affected by minor opacities from CO, $H_2S$, OCS, and $CH_4$, though their spectral features cannot be robustly detected in the spectrum. We show a model without CO and $CH_4$ in a) to demonstrate this, with the minor contribution by CO additionally highlighted in e).

# Methods

## Data Reduction

We produced multiple analyses of stellar spectra from the Stage 1 2D spectral images produced using the default STScI JWST Calibration Pipeline[45] ("rateints" files) and via customised runs of the STScI JWST Calibration Pipeline with user-defined inputs and processes for steps such as the "jump detection" and "bias subtraction" steps.

Each pipeline starts with the raw "uncal" 2D images which contain group-level products. As we noticed that the default superbias images were of poor quality, we produced two customised runs of the JWST Calibration Pipeline, using either the default bias step, or a customised version. The customised step built a pseudo-bias image by computing the median pixel value in the first group across all integrations, then subtracted the new bias image from all groups. We note that the poor quality of the default superbias images impacts NRS1 more significantly than NRS2, and this method could be updated once a better superbias is available.

Prior to ramp fitting, both our standard and custom bias step runs of the edited JWST Calibration Pipeline "destriped" the group-level images to remove so-called "1/*f* noise" (correlated noise arising from the electronics of the readout pattern, which appears as column striping in the subarray images[11,12]). Our group-level destriping step used a mask of the trace 15σ from the dispersion axis for all groups within an integration, ensuring a consistent set of pixels is masked within a ramp. The median values of non-masked pixels in each column were then computed and subtracted for each group.

The results of our customised runs of the JWST Calibration Pipeline are a set of custom group-level destriped products and custom bias-subtracted group-level destriped products. In both cases, the ramp-jump detection threshold of the JWST Calibration Pipeline was set to 15σ (as opposed to the default of 4σ) as it produced the most consistent results at the integration level. In both custom runs of the JWST Calibration Pipeline, all other steps and inputs were left at the default values.

For all analyses, wavelength maps from the JWST Calibration Pipeline were used to produce wavelength solutions, verified against stellar absorption lines, for both detectors. The mid-integration times in $BJD_{TDB}$ were extracted from the image headers for use in producing light curves. None of our data-reduction pipelines performed a flat-field correction since the available flat fields were of poor quality and unexpectedly removed portions of the spectral trace. In general, we found that 1/*f* noise can be corrected at either the group or integration level to similar effect; however, correction at the group-level with a repeated column-by-column cleaning step at the integration level likely results in cleaner 1D stellar spectra. This was particularly true for NRS2, due to the limited number of columns in which the unilluminated region on the detector extends both above and below the spectral trace, as shown in Extended Data Figure 1.

Below we detail each of the independent data reduction pipelines used to produce the time series of stellar spectra from our G395H observations.

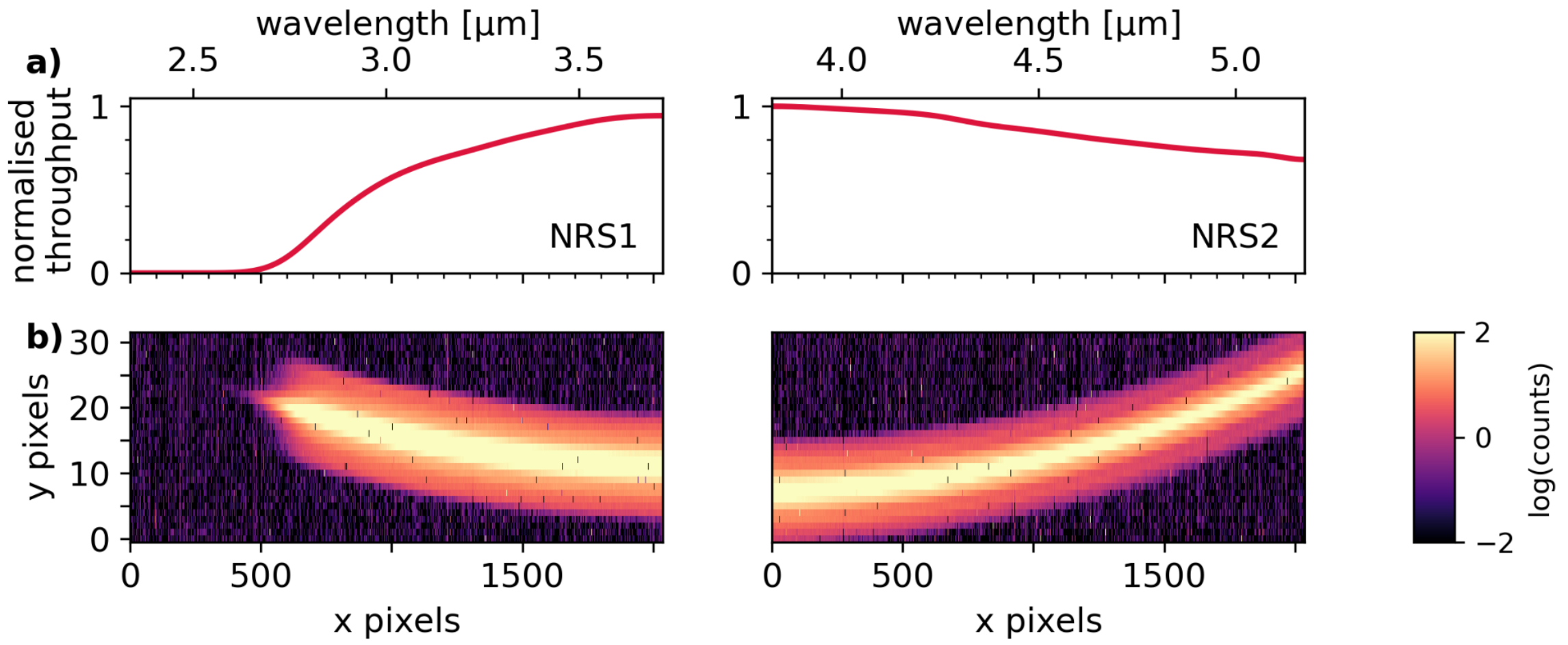

**Extended Data Figure 1: The throughput and spectral trace for WASP-39 across NRS1 and NRS2.** a) normalised throughput of NRS1 and NRS2 detectors (as custom produced, see Methods: Limb-darkening), which shows the cut-off at short wavelengths. b) 2D spectral images of the trace produced from the ExoTiC-JEDI [V1] reduction prior to cleaning steps. The aspect ratio has been stretched in the y direction to show the structure of the trace over the 32-pixel-wide subarray more clearly. The NRS2 spectral position is slightly offset from that of NRS1, as the NRS2 subarray was moved following commissioning to ensure that the centre of the spectral trace fell fully on the detector and did not fall off the top right corner[46].

**ExoTiC-JEDI pipeline**

We used the Exoplanet Timeseries Characterisation - JWST Extraction and Diagnostics Investigator (ExoTiC-JEDI[47]) pipeline on our custom group-level destriped products, treating each detector separately. Using the data-quality flags produced by the JWST Calibration Pipeline, we replaced any pixels identified as bad, saturated, dead, hot, low quantum efficiency or no gain value with the median value of surrounding pixels. We additionally searched each integration for pixels that were spatial outliers from the median of the surrounding 20 pixels in the same row by 6σ (to remove permanently affected "bad" pixels), or outliers from the median of that pixel in the surrounding 10 integrations in time by 20σ (to identify high energy short-term effects such as cosmic rays), and replaced the outliers with the median values. To obtain the trace position and full-width-at-half-maximum (FWHM), we fitted a Gaussian to each column of an integration, finding a median standard deviation of 0.7 pixels. A 4th-order polynomial was fitted through the trace centres and the widths, which were smoothed with a median filter, to obtain a simple aperture region. This region extended five times the FWHM of the spectral trace, above and below the centre, corresponding to a median aperture width of 7 pixels. To remove any remaining $1/f$ and background noise from each integration, we subtracted the median of the unilluminated region in each column by masking all pixels 5 pixels away from the aperture. For each integration, the counts within each row and column of the aperture region were summed using an intrapixel extraction, taking the relevant fractional flux of the pixels at the edge of the aperture, and cross-correlated to produce x- and y-pixel shifts for detrending (see Extended Data Figure 2). On average the x-pixel shift represents movement of $1\times10^{-4}$ and $8\times10^{-6}$ of a pixel for NRS1 and NRS2 respectively. The aperture column sums resulted in 1D stellar spectra with errors calculated from photon noise after converting from data numbers using the gain factor. This reduction is denoted hereafter as ExoTiC-JEDI [V1].

We produced additional 1D stellar spectra from both the custom group-level destriped product and custom bias-subtracted group-level destriped products using the ExoTiC-JEDI pipeline as described above, but with additional cleaning by repeating the spatial outliers step. The reduction with additional cleaning using the custom group-level destriped products is hence denoted as ExoTiC-JEDI [V2], and the reduction with additional cleaning using the custom bias-subtracted group-level destriped products is hence denoted as ExoTiC-JEDI [V3].

**Tiberius pipeline**

We used the Tiberius pipeline, which builds upon the LRG-BEASTS spectral reduction and analysis pipelines[15,49,50], on our custom group-level destriped products. For each detector, we created bad-pixel masks by manually identifying hot pixels within the data. We then combined them with pixels flagged as greater than 3σ above the defined background. Prior to identifying the spectral trace, we interpolated each column of the detectors onto a grid 10× finer than the initial spatial resolution. This step reduces the noise in the extracted data by improving the extraction of flux at the sub-pixel level, particularly where the edges of the photometric aperture bisect a pixel. We also interpolated over the bad pixels using their nearest neighbouring pixels in x and y.

We traced the spectra by fitting Gaussians at each column and used a running median, calculated with a moving box with a width of five data points, to smooth the measured centres of the trace. We fitted these smoothed centres with a 4th-order polynomial, removed points that deviated from the median by 3σ, and refitted with a 4th-order polynomial. To remove any residual background flux not captured by the group-level destriping, we fitted a linear polynomial along each column, masking the stellar spectrum. This was defined by an aperture of a width of four pixels centred on the trace. We also masked an additional 7 pixels on either side of the aperture so that the background was not fitting the wings of the stellar PSF, and we clipped any pixels within the background that deviated by more than 3σ from the mean for that particular column and frame. After removing the background in each column, the stellar spectra were then extracted by summing within a 4-pixel wide aperture and correcting for pixel oversampling caused by the interpolation onto a finer grid, as described above. The uncertainties in the stellar spectra were calculated from the photon noise prior to background subtraction.

**transitspectroscopy pipeline**

We used the transitspectroscopy pipeline[51] on the "rateints" products of the JWST Calibration Pipeline, treating each detector separately. The trace position was found from the median integration by cross-correlating each column with a Gaussian function, removing outliers using a median filter with a 10-pixel-wide window, and smoothing the trace with a spline. We removed $1/f$ noise from the "rateints" products by masking all pixels within 10 pixels from the centre of the trace, and calculating and removing the median value from all columns. We then used optimal extraction[52] to obtain the 1D stellar spectra, with a 5-pixel-wide aperture above and below the trace. This allowed us to treat bad pixels and cosmic rays that had not been accounted for or masked in the "rateints" products in an automated fashion. To monitor systematic trends in the observations, we also calculated the trace centre as described above, and the FWHM for all integrations. The FWHM was calculated at each column and at each integration by first subtracting each column to half the maximum value in it, with a spline used to interpolate the profile. The roots of this profile were then found in order to estimate the FWHM.

### Eureka! pipeline

We used two customised versions of the Eureka! pipeline[53] which combines standard steps from the JWST Calibration Pipeline with an optimal extraction scheme in order to obtain the time series of stellar spectra.

The first Eureka! reduction used our custom group-level destriped products, and applied Stages 2 and 3 of Eureka!. Stage 2, a wrapper of the JWST Calibration Pipeline, followed the default settings up to the flat fielding and photometric calibration steps, which were both skipped. Stage 3 of Eureka! was then used to perform the background subtraction and extraction of the 1D stellar spectra. We started by correcting for the curvature of NIRSpec G395H spectra by shifting the detector columns by whole pixels, to bring the peak of the distribution of the counts in each column to the centre of our subarray. To ensure that this curvature correction was smooth, we computed the shifts in each column for each integration from the median integration frame in each segment, and applied a running median to the shifts obtained for each column. The pixel shifts were applied with periodic boundary conditions, such that pixels shifted upwards from the top of the subarray appeared at the bottom after the correction, ensuring no pixels were lost. We applied a column-by-column background subtraction by fitting and subtracting a flat line to each column of the curvature-corrected data frames, obtained by fitting all pixels further than 6 pixels from the central row. We also performed a double iteration of outlier rejection in time with a threshold of 10σ, along with a 3σ spatial outlier rejection routine, to ensure bad pixels were not biasing our background correction. These outlier rejection thresholds were selected to remove clear outliers in the data and provide a balance with the background subtraction step. We performed optimal extraction using an extraction profile defined from the median frame, the central 9 rows of our subarray (4 rows on either side of the central row). We also measured the vertical shift in pixels of the spectrum from one integration to the other using cross-correlation, and the average point spread function width at each integration, obtained by adding all columns together and fitting a Gaussian to the profile to estimate its width. This reduction is henceforth denoted as Eureka! [V1].

The second Eureka! reduction (Eureka! [V2]) used the "rateints" outputs of the JWST calibration pipeline, and applied Stage 2 of Eureka! as described above, with a modified version of Stage 3. In this reduction, we corrected the curvature of the trace using a spline and found the centre of the trace using the median of each column. We removed $1/f$ noise by subtracting the mean from each column, excluding the region 6 pixels away from the trace, sigma clipping outliers at 3σ. We extracted the 1D stellar spectra using a 4-pixel wide aperture on either side of the trace centre.

## Limb-darkening

Limb-darkening is a function of the physical structure of the star that results in variations in the specific intensity of the light from the centre of the star to the limb such that the limb looks darker than the centre. This is due to the change in depth of the stellar atmosphere being observed. At the limb of the star, the region of the atmosphere being observed at slant geometry is at higher altitudes and lower density and thus lower temperatures, compared to the deeper atmosphere observed at the centre of the star where hotter denser layers are observed. The effect of limb-darkening is most prominent at shorter wavelengths, resulting in a more U-shaped light curve compared to the flat-bottomed light curves observed at longer wavelengths. To account for the effects of limb-darkening on the time series light curves, we used analytical approximations for computing the ratio of the mean intensity to the central intensity of the star. The most commonly used limb-darkening laws for exoplanet transit light curves are the quadratic, square-root, and non-linear 4-parameter laws[54]:

Quadratic:

$$\frac{I(\mu)}{I(1)} = 1 - u_1(1 - \mu) - u_2(1 - \mu)^2$$

Square-root:

$$\frac{I(\mu)}{I(1)} = 1 - s_1(1 - \mu) - s_2\left(1 - \sqrt{\mu}\right)$$

Non-linear 4-parameter:

$$\frac{I(\mu)}{I(1)} = 1 - c_1\left(1 - \mu^{0.5}\right) - c_2(1 - \mu) - c_3\left(1 - \mu^{1.5}\right) - c_4\left(1 - \mu^2\right)$$

where $I(1)$ is the specific intensity in the centre of the disk, $u_1$, $u_2$, $s_1$, $s_2$, $c_1$, $c_2$, $c_3$, $c_4$ are the limb-darkening coefficients, and $\mu = \cos(\gamma)$, where $\gamma$ is the angle between the line of sight and the emergent intensity.

The limb-darkening coefficients depend on the particular stellar atmosphere and therefore vary from star to star. For consistency across all of the light curve fitting, we used 3D stellar models[55] for $T_{\rm eff}$ = 5512 K, log($g$) = 4.47 cgs, and [Fe/H] = 0.0, along with the instrument throughput to determine $I$ and $\mu$. As instrument commissioning showed that the throughput was higher than the pre-mission expectations[56], a custom throughput was produced from the median of the measured ExoTiC-JEDI [V2] stellar spectra, divided by the stellar model, and Gaussian smoothed.

Where the limb-darkening coefficients were held fixed, we used the values computed using the ExoTiC-LD[57,58] package which can compute the linear, quadratic, and 3- and 4-parameter non-linear limb-darkening coefficients[54,59]. To compute and fit for the coefficients from the square-root law, we used previously outlined formalisms[60,61]. We highlight that we do not see any dependence in our transmission spectra on the limb-darkening procedure used across our independent reductions and analyses.

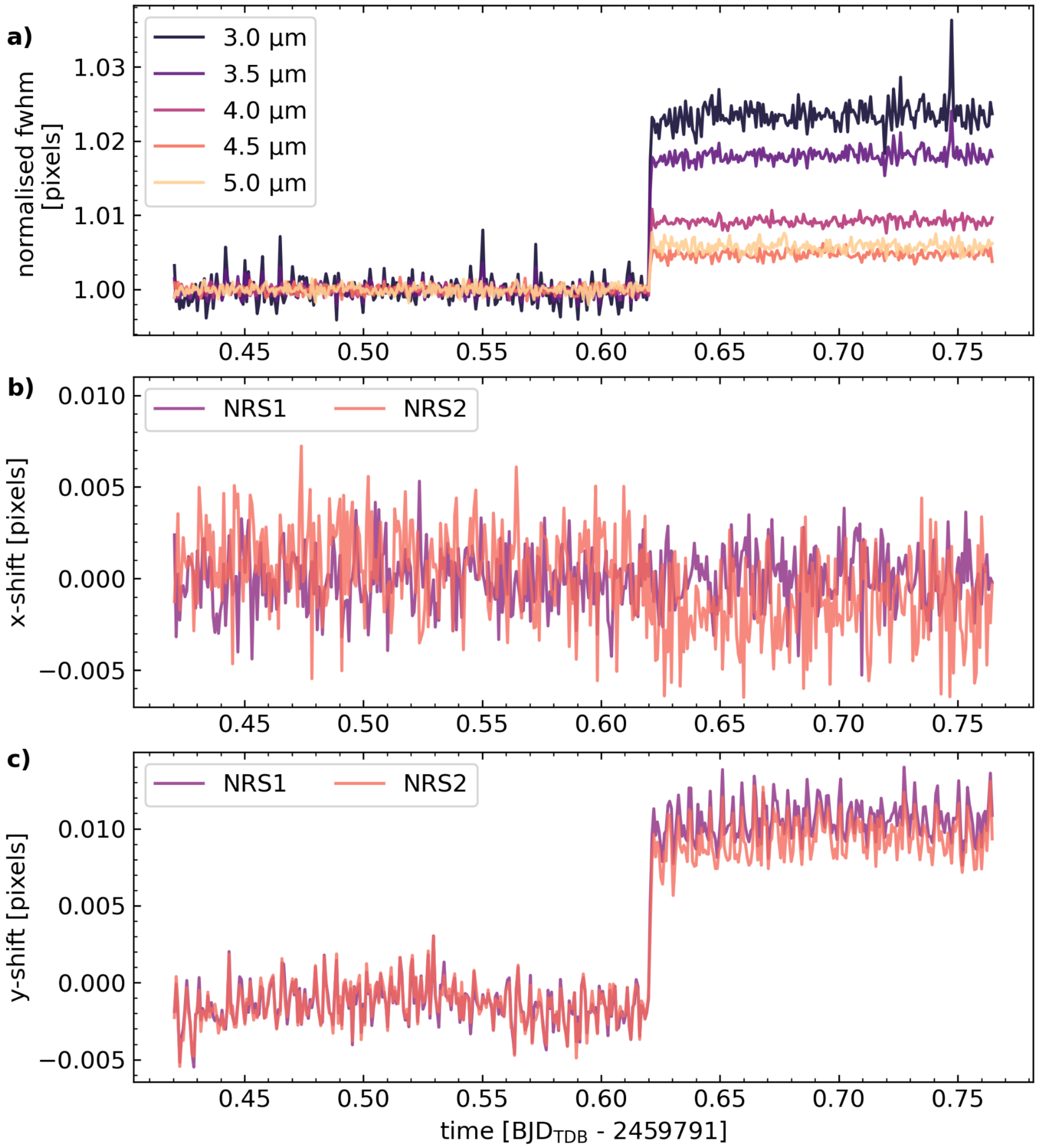

**Extended Data Figure 2: Time-dependent decorrelation parameters.** a) the change in the FWHM of the spectral trace at selected wavelengths. This change does not correspond to any high gain antenna moves and is attributed to a large mirror tilt event. These measurements demonstrate that the mirror tilt event has a wavelength dependence. Changes to the PSF have a larger impact at short wavelengths, as the PSF of the spectrum increases with wavelength[46]. b) and c) the change in the x- and y-pixel position of the spectral trace as functions of time respectively. Positional shifts are calculated by cross-correlating the spectral trace with a template to measure sub-pixel movement on the detector. The y position shift clearly shows a link to the mirror tilt event.

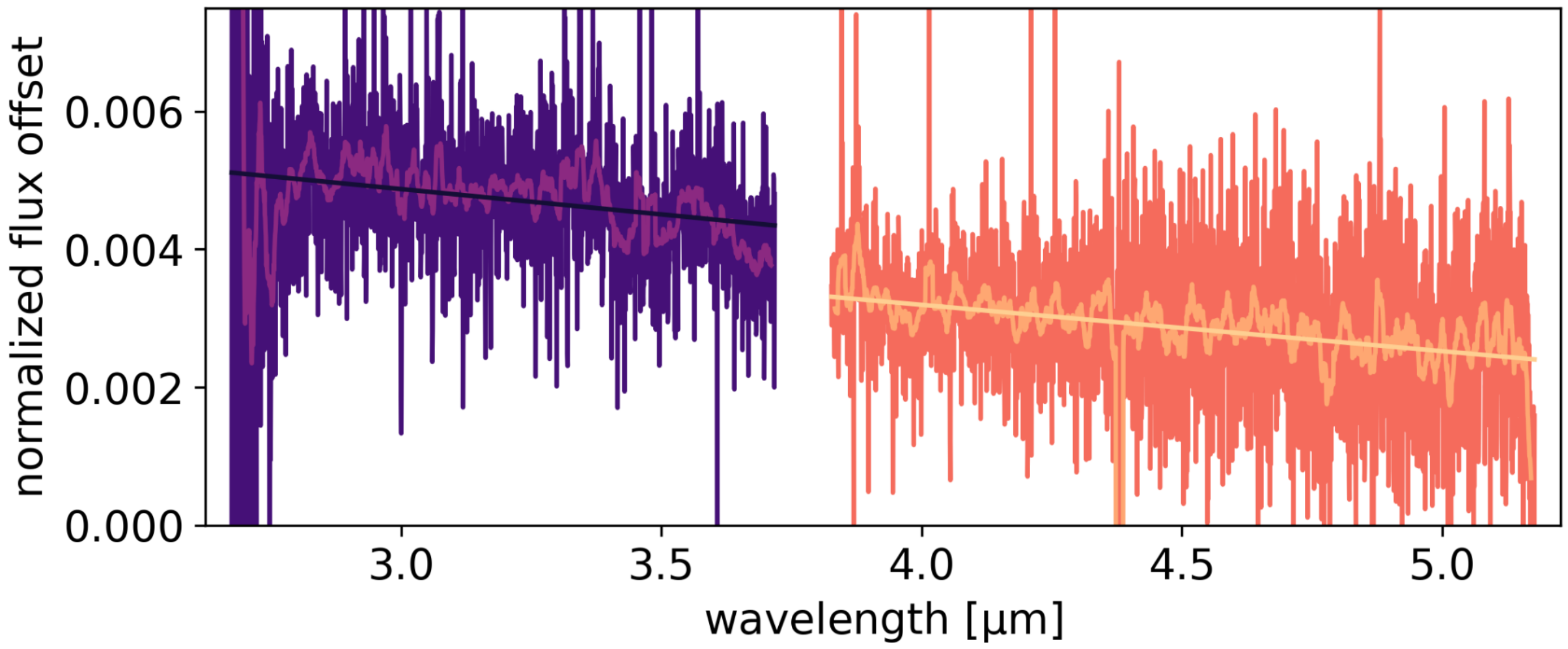

**Extended Data Figure 3: Normalised flux offset of the stellar baseline before and after the tilt event as a function of wavelength for NRS1 (purple) and NRS2 (orange).** The normalised flux offset is calculated per pixel by measuring the median flux in the stellar baseline before and after the transit and calculating the difference. These differences are then normalised by the before-transit flux and plotted on a common scale. Overplotted are the data binned to a resolution of 10 pixels to match our presented transmission spectra (Figure 2). We also show a linear fit to each detector to better quantify the decreasing tilt flux amplitude with increasing wavelength (NRS1 = -0.00073374$x$ + 0.00707344, NRS2 = -0.00067165$x$ + 0.00588128).

**Light Curve Fitting**

From the time series of extracted 1D stellar spectra, we created our broadband transit light curves by summing the flux over 2.725–3.716 µm for NRS1 and 3.829–5.172 µm for NRS2. For the spectroscopic light curves, we used a common 10-pixel binning scheme within these wavelength ranges to generate a total of 349 spectroscopic bins (146 for NRS1 and 203 for NRS2). We also tested wider and narrower binning schemes but found that 10-pixel wide bins achieved the best compromise between the noise in the spectrum and showcasing the abilities of G395H across multiple analyses. In our analyses, we treated the NRS1 and NRS2 light curves independently to account for differing systematics across the two detectors. To construct the full NIRSpec G395H transmission spectrum of WASP-39b, we fitted the NRS1 and NRS2 broadband and spectroscopic light curves using 11 independent light curve fitting codes, which are detailed below. Where starting values were required, all analyses used the same system paramters[38]. In many of our analyses, we detrended the raw broadband and spectroscopic light curves using the time-dependent decorrelation parameters for the change in the FWHM of the spectral trace or the shift in x- and y-pixel positions (Extended Data Figure 2). We also used various approaches to account for the mirror tilt event, which we found to have a smaller effect at longer wavelengths (Extended Data Figure 3).

Using fitting pipeline 1 we measured a centre of transit time ($T_0$) of $T_0$ = 2459791.612039 ± 0.000017 $BJD_{TDB}$ and $T_0$ = 2459791.6120689 ± 0.000021 computed from the NRS1 and NRS2 broadband light curves respectively; a majority of fitting pipelines obtained $T_0$ within 1σ of the quoted uncertainty.

For each of our analyses, we computed the expected photon noise from the raw counts taking into account the instrument read noise (16.18 $e^-$ on NRS1 and 17.75 $e^-$ on NRS2), gain (1.42 for NRS1 and 1.62 for NRS2), and the background counts (which are found to be negligible after cleaning) and compared it to the final signal-to-noise (S/N) in our light curves (see Main Text Figure 1). We also determine the level of white and red noise in our spectroscopic light curves by computing the Allan deviation[62] which is used to measure the deviation from the expected photon noise by binning the data into successively smaller bins (i.e., fewer data points per bin) and calculating the S/N achieved[63]. Extended Data Figure 4 shows the Allan deviation for 3 of the 11 reductions performed on the data (see the ExoTiC-ISM noise_calculator function[57]).

While there is a general consensus across each of the data analyses, by comparing the results of each fitting pipeline we were better able to evaluate the impact of different approaches to the data reduction, such as the removal of bad pixels. For future studies, we recommend the application of multiple pipelines that utilise differing analysis methods, such as the treatment of limb-darkening, systematic effects, and noise removal. No single pipeline presented on its own can fully evaluate the measured impact of each effect, given the differing strategies, targets, and potential for chance events such as a mirror tilt with each observation. In particular, attention should be paid to 1/f noise removal at the group- versus integration-level for observations with fewer groups per integration than this study.

Below we detail each of our 11 fitting pipelines, and summarise them in Extended Data Table 1.

**Fitting pipeline 1 - ExoTiC-JEDI**

We fitted the broadband and spectroscopic light curves produced from the ExoTIC-JEDI [V3] stellar spectra using the least-squares optimizer, scipy.optimize lm[64]. We simultaneously fitted a batman transit model[65] with a constant baseline and systematics models for data pre- and post-tilt event, fixing the centre of transit time $T_0$, the ratio of the semi-major axis to stellar radius $a/R_\star$, and the inclination $i$, to the broadband light curve best-fit values. The systematics models included a linear regression on x and y, where x and y are the measured trace positions in the dispersion and cross-dispersion directions, respectively. We accounted for the tilt event by normalising the light curve pre-tilt by the median pre-transit flux and normalising the light curve post-tilt by the median post-transit flux. We discarded the first 15 integrations and the 3 integrations during the tilt event. 14-pixel columns were discarded due to outlier pixels directly on the trace. We fixed the limb-darkening coefficients to the 4-parameter nonlinear law.

**Fitting pipeline 2 - Tiberius**

We used the broadband light curves generated from the Tiberius stellar spectra and fitted for the ratio of the planet to stellar radii $R_p/R_\star$, as well as $i$, $T_0$, $a/R_\star$, the quadratic law limb-darkening coefficient $u_1$, and the systematics model parameters, the x- and y-pixel shifts, FWHM, and sky background, with the period $P$, the eccentricity $e$, and $u_2$ fixed. We used uniform priors for all the fitted parameters. Our analytic transit light curve model was generated with batman. We fitted our broadband light curve with a transit+systematics model using a Gaussian Process (GP)[66,67], implemented through george[68], and a Markov Chain Monte Carlo (MCMC), implemented through emcee[69]. For our Tiberius spectroscopic light curves, we held $a/R_\star$, $i$, and $T_0$ fixed to the values determined from the broadband light curve fits, and applied a systematics correction from the broadband light curve fit to aid in fitting the mirror tilt event. We fitted the spectroscopic light curves using a GP with an exponential squared kernel for the same systematics detrending parameters detailed above. We used a Gaussian prior for $a/R_\star$ and uniform priors for all other fitted parameters.

**Fitting pipeline 3 - Aesop**

We used transit light curves from the ExoTiC-JEDI [V1] stellar spectra, and fit the broadband and spectroscopic light curves using the least-squares minimizer LMFIT[70]. We fitted each light curve with a two-component function consisting of a transit model (generated using batman) multiplied by a systematics model. Our systematics model included the x- and y-pixel positions on the detector (x, y, xy, $x^2$, and $y^2$). To capture the amplitude of the tilt event in our systematics model, we did a piecewise linear regression to the out-of-transit baseline pre- and post-tilt. We first fit the broadband light curve by fixing $P$ and $e$, and fitting for $T_0$, $a/R_\star$, $i$, $R_p/R_\star$, stellar baseline flux, and systematic trends using wide uniform priors. For the spectroscopic light curves, we fixed $T_0$, $a/R_\star$, and $i$ to the best-fit values from the broadband light curve and fit for $R_p/R_\star$. We held the non-linear limb-darkening coefficients fixed.

**Fitting pipeline 4 - transitspectroscopy**

We fit the broadband and spectroscopic light curves produced from the transitspectroscopy stellar spectra, running juliet[71] in parallel with the light curve fitting module of the transitspectroscopy pipeline[51] with dynamic nested sampling via dynesty[72] and analytical transit models computed using batman. We fit the broadband light curves for NRS1 and NRS2 individually, fixing $P$, $e$, and ω, and fitting for the impact parameter $b$, as well as $T_0$, $a/R_\star$, $R_p/R_\star$, extra jitter, and the mean out-of-transit flux. We also fitted two linear regressors, a simple slope and a ‘jump’ (a regressor with zeros prior to the tilt event and ones after the tilt event), scaled to fit the data. We fitted the square-root law limb-darkening coefficients using the Kipping sampling scheme. We fitted the spectroscopic light curves at the native resolution of the instrument, fixing $T_0$, $a/R_\star$, and $b$. We used the broadband light curve systematics model for the spectroscopic light curve, with wide uniform priors for each wavelength bin, and set truncated normal priors for the square-root law limb-darkening coefficients. We also fitted a jitter term added in quadrature to the error bars at each wavelength with a log-uniform prior between 10 and 1000 ppm. We computed the mean of the limb-darkening coefficients by first computing the non-linear coefficients from ATLAS models[73] and passing them through the SPAM algorithm[74]. We binned the data into 10-pixel wavelength bins after fitting the native resolution light curves.

**Fitting pipeline 5 - ExoTEP**

We fitted the transit light curves from the Eureka! [V1] stellar spectra using the ExoTEP analysis framework[75–78]. ExoTEP uses batman to generate analytical light curve models, adds an analytical instrument systematics model along with a photometric scatter parameter, and fits for the best-fit parameters and their uncertainties using emcee. Prior to fitting, we cleaned the light curves by running 10 iterations of 5σ clipping using a running median of window length 20 on the flux, x- and y-pixel shifts and the ‘ydriftwidth’ data product from Eureka! Stage 3 (the average spatial PSF width at each integration). Our systematics model consisted of a linear trend in time with a ‘jump’ (constant offset) after the tilt event. The 'ydriftwidth' was used prior to the fit in order to locate the tilt event. We used a running median of 'ydriftwidth' to search for the largest offset, and flagged every data point after the tilt event so that they would receive a constant 'jump' offset in our systematics model. We also removed the first point of the tilt event in our fits since it was not captured by the "jump" model. We fitted the broadband light curves, fitting for $R_p/R_\star$, photometric scatter, $T_0$, $b$, $a/R_\star$, the quadratic limb-darkening coefficients, and the systematics model parameters (normalisation constant, slope in time and constant 'jump' offset). We used uninformative flat priors on all the parameters. The orbital parameters were fixed to the best-fit broadband light curve values for the subsequent spectroscopic light curve fits.

**Fitting pipeline 6**

We fitted transit light curves from the ExoTiC-JEDI [V1] stellar spectra using a custom lmfit light curve fitting code. The final systematic detrending model included a batman analytical transit model multiplied by a systematics model consisting of a linear stellar baseline term, a linear term for the x- and y-pixel shifts, and an exponential ramp function. The tilt event was accounted for by decorrelating the light curves with the y-pixel shifts, using a (1 + constant × y-shift) term with the constant fitted for in each light curve. For the broadband light curve fits, we fixed $P$ and fitted for $T_0$, $i$, $R_p/R_\star$, $a/R_\star$, x- and y-pixel shifts, and the exponential ramp amplitude and timescale. We fixed the non-linear limb-darkening coefficients. For the spectroscopic light curve fits, we fixed the values of $T_0$, $i$, and $a/R_\star$, and the exponential ramp timescale to the broadband light curve fit values, and fitted for $R_p/R_\star$, the x- and y-shifts, and the ramp amplitude. Wide, uniform priors were used on all the fitting parameters in both the broadband and spectroscopic light curve fits.

**Fitting pipeline 7**

We fitted transit light curves from the Eureka! [V2] stellar spectra, using Pylightcurve[78] to generate the transit model with emcee as the sampler. We calculated the non-linear 4-parameter limb-darkening coefficients using ExoTHETyS[79], which relies on PHOENIX 2012-2013 stellar models[80,81], and fixed these in our fits to the precomputed theoretical values. Our full transit+systematics model included a transit model multiplied by a 2nd-order polynomial in the time domain. We accounted for the tilt event by subtracting the mean of the last 30 integrations of the pre-transit data from the mean of the first 30 integrations of the post-transit data, to account for the jump in flux, shifting the post-transit light curve upwards by the jump value. We fitted for the systematics (the parameters of the 2nd-order polynomial), $R_p/R_\star$, and $T_0$. We used uniform priors for all the fitted parameters. We adopted the RMS of the out-of-transit data as the error bars for the light curve data points to account for the scatter in the data.

**Fitting pipeline 8**

We used the transit light curves generated from the ExoTiC-JEDI [V1] stellar spectra. We fit the broadband light curves with a batman transit model multiplied by a 2nd-order systematics model as a function of x- and y-pixel shifts. We fixed both the quadratic limb-darkening coefficients for each wavelength bin. We fitted for $R_p/R_\star$, $i$, $T_0$ and $a/R_\star$, using wide uninformed priors, and ran our fits using emcee. For the spectroscopic light curve fits, we fixed $i$ and $a/R_\star$ to the broadband light curve best-fit values, and fitted for an additional error term added in quadrature.

**Fitting pipeline 9**

We used the transit light curves from the ExoTiC-JEDI [V1] stellar spectra. We fixed both the quadratic limb-darkening coefficients, and fitted the light curves with a batman transit model multiplied by a systematics model of a 2nd-order function of x- and y-pixel shifts. We fixed the best-fit broadband light curve values for $T_0$, $a/R_\star$, and $i$ for the spectroscopic light curve fits, and fitted for $R_p/R_\star$ using emcee for each 10-pixel bin, with the walkers initialised in a tight cluster around the best-fit solution from a Levenberg-Marquardt minimisation. For both the broadband and spectroscopic light curves, we also fit for an additional per-point error term.

**Fitting pipeline 10**

We fitted the transit light curves from the ExoTiC-JEDI [V2] stellar spectra and performed our model fitting using automatic differentiation implemented with JAX[82]. We used a GP systematics model with a time-dependent Matérn ($\nu = 3/2$) kernel and a variable white noise jitter term. The mean function was comprised of a linear trend in time plus a sigmoid function to account for the drop in measured flux that occurred mid-transit due to the mirror tilt event. For the transit model, we used the exoplanet package[83], making use of previously developed light curve models[84,85]. For the GP systematics component, a generalisation of the algorithm used by the celerite package[86] was adapted for JAX. We fixed both the quadratic limb-darkening coefficients. For the initial broadband light curve fit, both NRS1 and NRS2 were fitted simultaneously. All transit parameters were shared across both light curves, except for $R_p/R_\star$ which was allowed to vary for NRS1 and NRS2 independently. We fitted for $T_0$, the transit duration, $b$, and both $R_p/R_\star$ values. For the spectroscopic light curve fits, all transit parameters were then fixed to the maximum-likelihood values determined from the broadband fit, except for $R_p/R_\star$, which was allowed to vary for each wavelength bin. Uncertainties for the transit model parameters, including $R_p/R_\star$, were assumed to be Gaussian and estimated using the Fisher information matrix at the location of the maximum likelihood solution, which was computed exactly using the JAX automatic differentiation framework.

**Fitting pipeline 11 - Eureka!**

We used transit light curves from the Eureka! [V2] time series stellar spectra with the open-source Eureka! code to estimate the best-fit transit parameters and their uncertainties using a Markov Chain Monte Carlo (MCMC) fit to the data (implemented by emcee). A linear trend in time was used as a systematic model and a step function was used to account for the tilt event. We fixed $a/R_\star$, $i$, $T_0$, and the time of the tilt event to the best-fit values from the NRS1 broadband light curve, with the three integrations during the tilt event clipped from the light curve. We fitted for $R_p/R_\star$, both quadratic limb-darkening coefficients, the linear time trend, and the magnitude of the step from the tilt event, with uniform priors for both the magnitude of the step and the limb-darkening coefficients.

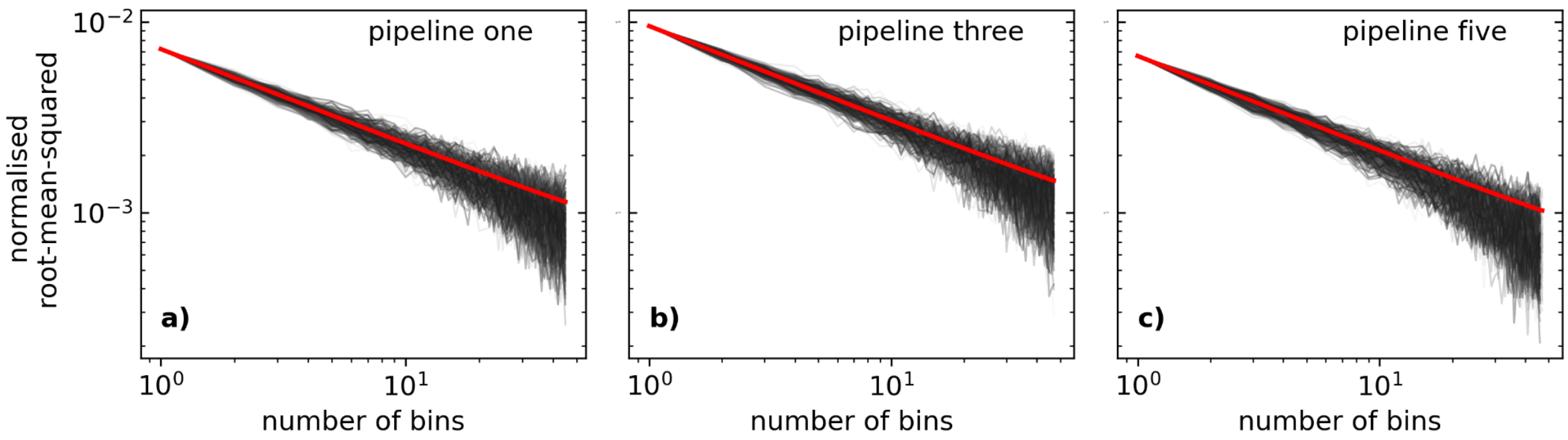

**Extended Data Figure 4: Normalised root-mean-squared binning statistic for three of the 11 reductions detailed in Methods.** In each subplot, the red line shows the expected relationship for perfect Gaussian white noise. The black lines show the observed noise from each spectroscopic light curve for pipelines 1, 3, and 5. In order to compare bins and noise levels, values for all bins in each pipeline are normalised by dividing by the value for a bin width of 1.

**Transmission spectral analysis**

Based on the independent light curve fits described above, we produced 11 transmission spectra from our NIRSpec G395H observations using multiple analyses and fitting methods. Extended Data Table 1 shows a breakdown of the different steps used in each fitting pipeline. In this work, 3 different 2D spectral image products were used, producing 7 different 1D stellar spectra. 11 different fitting pipelines using 5 different limb-darkening methods were then applied. Each of these fitting pipelines resulted in an independent analysis of the observations and 11 comparative transmission spectra. Extended Data Figure 5 details comparative information for all 11 analyses to quantify their similarities and differences.

We computed the standard deviation of the 11 spectra in each wavelength bin and compared this to the mean uncertainty obtained in that bin. The average standard deviation in each bin across all fitting pipelines was 199 ppm, compared to an average uncertainty of 221 ppm (which ranged from 131–625 ppm across the bins). The computed standard deviation in each bin across all pipelines ranged from 85–1040 ppm, with greater than 98% of the bins having a standard deviation lower than 500 ppm. We see an increase in scatter at longer wavelengths, with the structure of the scatter following closely with the measured stellar flux, where throughput beyond 3.8 μm combines with decreasing stellar flux. The unweighted mean uncertainty of all 11 transmission spectra follows a similar structure to the standard deviation, with the uncertainty increasing at longer wavelengths. The uncertainties from each fitting pipeline are consistent to within 3σ of each other, with the uncertainty per bin typically overestimated compared to the mean uncertainty across all reductions.

From all 11 transmission spectra, we computed a weighted average transmission spectrum using the transit depth values from all reductions in each bin weighted by 1/variance ($1/\sigma^2$, where σ is the uncertainty on the data point from each reduction). For this weighted average transmission spectrum, the unweighted mean of the uncertainties in each bin was used to represent the errorbar on each point. By using the weighted average of all 11 independently obtained transmission spectra, we therefore do not apply infinite weight to any one reduction in our interpretation of the atmosphere. While the weighted average could be sensitive to any one spectrum with underestimated uncertainties, we find that our uncertainties on average are overestimated compared to the average. Similarly, we chose to use the mean rather than the median of the transmission spectral uncertainties as this results in a more conservative estimate of the uncertainties in each bin. We find that all of the 11 transmission spectra are within 2.95σ of the weighted average transmission spectrum without applying offsets.

We calculated normalised transmission spectrum residuals for each fitting pipeline by subtracting the weighted average spectrum and dividing by the uncertainty in each bin. We generated histograms of the normalised transmission spectrum residuals and used the mean and standard deviation of the residuals to compute a normalised probability density function (PDF). We performed a K-S test on each of the normalised residuals and found that all are approximately symmetric around their means, with normal distributions. This confirms that they are Gaussian in shape, with the null hypothesis that they are not Gaussian strongly rejected in a majority of cases (see Extended Data Figure 5).

The PDFs of the residuals indicate three distinct clusters of computed spectra based on their deviations from the mean, and their spreads. The first cluster is negatively offset by less than 200 ppm and corresponds to fitting pipelines which used extracted stellar spectra which underwent additional cleaning steps (e.g., ExoTiC-JEDI [V3]). The second cluster is positively offset from the mean by ~120 ppm, and contains the majority of the transmission spectra produced. We see no obvious trends within this group to any specific reduction or fitting process. The final cluster is centred around the mean, but has a broad distribution, suggesting a larger scatter both above and below the average transmission spectrum. This is likely the result of uncleaned outliers or marginal offsets between NRS1 and NRS2. These transmission spectra demonstrate that the 11 independent fitting pipelines are able to accurately reproduce the same transmission spectral feature structures, further highlighting the minimised impact of systematics on the time series light curves. We suspect that the minor differences resulting from different reduction products and fitting pipelines are linked to the super-bias and treatment of 1/f noise. We anticipate that the impacts of these will be improved with new super-bias images, expected to soon be released by STScI, and with more detailed investigation into the impact of 1/f noise at the group level beyond the scope of this work.

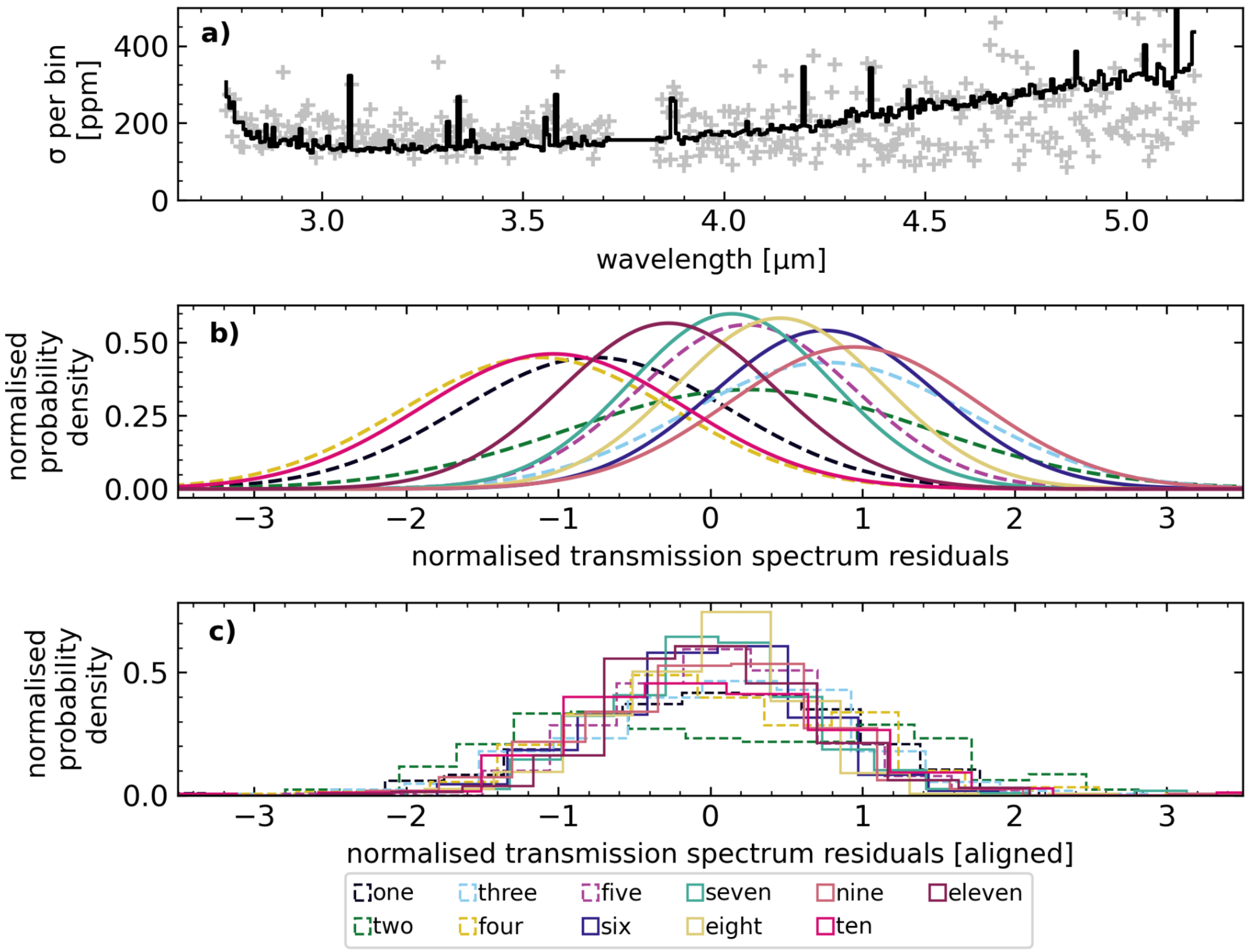

**Extended Data Figure 5: Comparison between all fitting pipelines performed on the spectroscopic light curves.** a) the underlying grey data points show the standard deviation between all transmission spectra per spectral bin. The black line shows the unweighted mean uncertainty on the transit depth per bin. Spikes in the uncertainties correspond to spectral bins with higher standard deviations, likely due to differences in pixel flagging or sigma-clipping at the light curve level. b) Gaussian probability density functions (PDF) of the normalised transmission spectrum residuals, showing the mean offset and the spread relative to the weighted average transmission spectrum. c) histograms of the normalised transmission spectrum residuals aligned to zero by subtracting the mean of the distribution that was used to generate the PDF above. In panels b) and c), the coloured lines and numbers correspond to the fitting pipeline used to obtain each transmission spectrum, as summarised in Extended Data Table 1. The dashed lines correspond to the fitting pipeline results presented in Figure 2, demonstrating that they are drawn from across the distribution.

**Extended Data Table 1:** An outline of the combined products and fitting pipelines used to compute each transmission spectrum.

| Transmission spectrum | Spectral images | Stellar spectrum | Fitting process | Limb-darkening |
|---|---|---|---|---|
| 1 | custom bias subtracted group-level destriped | ExoTiC-JEDI [V3] | LM, x, y, tilt normalised | non-linear, fixed |
| 2 | custom group-level destriped | Tiberius | GP [squared exponential kernel] + MCMC | quadratic, fit u1, fixed u2 |
| 3 | custom group-level destriped | ExoTiC-JEDI [V1] | LM, x, y, linear regression | non-linear, fixed |
| 4 | default rateints | transitspectroscopy | Nested sampling, linear(t), step(t) | square-root, fit |
| 5 | custom group-level destriped | Eureka! [V1] | MCMC, linear(t), constant tilt step | quadratic, fit u1 & u2 |
| 6 | custom group-level destriped | ExoTiC-JEDI [V1] | LM, x, y, exponential ramp, tilt decorrelated against y-shifts | non-linear, fixed |
| 7 | default rateints | Eureka! [V2] | MCMC, quadratic(t), tilt normalised | non-linear (ExoTHETyS), fixed |
| 8 | custom group-level destriped | ExoTiC-JEDI [V1] | MCMC, x, y | quadratic, fixed |
| 9 | custom group-level destriped | ExoTiC-JEDI [V1] | MCMC, x, y | quadratic, fixed |
| 10 | custom group-level destriped | ExoTiC-JEDI [V2] | GP [time-dependent Matern kernel], linear(t), sigmoid function for tilt event | quadratic, fixed |
| 11 | default rateints | Eureka! [V2] | MCMC, linear(t), step(t) | quadratic, fit u1 & u2 |

MCMC - Markov chain Monte Carlo, LM - least-squares minimiser, GP - gaussian process

**Model Comparison**

To identify spectral absorption features, we compared the resulting weighted average transmission spectrum of WASP-39b to a number of 1D radiative-convective thermochemical equilibrium (RCTE) atmosphere models from three independent model grids. Each forward model is computed on a set of common physical parameters (e.g. metallicity, C/O ratio, internal temperature, and heat redistribution), shown in Extended Data Table 2. Additionally, each model grid contains different prescriptions for treating certain physical effects (e.g., scattering aerosols). While each grid contains different opacity sources from varying linelists (see Extended Data Table 2), they each consider all major molecular and atomic species[87]. Each model transmission spectrum from the grids was binned to the same resolution as that of the observations to compute the $\chi^2$ per data point, with a wavelength-independent transit depth offset as the free parameter. In general, the forward model grids fit the major features of the data, but are unable to place statistically significant constraints on many of the atmospheric parameters due to both the finite nature of the forward model grid spacing[13] and the insensitivity of some of these parameters to WASP-39b's 3-5 μm transmission spectrum (for example, >100K differences in interior temperature provided nearly identical $\chi^2$/N)

**ATMO**

We used the ATMO radiative-convective-thermochemical equilibrium grid[88–91], which consists of model transmission spectra for different day-night energy redistribution factors, atmospheric metallicities, C/O ratios, haze factors and grey cloud factors with a range of line lists and pressure-broadening sources[91]. In total, there were 5,160 models. Within this grid, we find the best-fit model to have 3x solar metallicity, with a C/O ratio of 0.35, and a grey cloud opacity 5x the strength of $H_2$ Rayleigh Scattering at 350 nm and a $\chi^2$/N = 1.098 for N = 344 data points and only fitting for an absolute altitude change in y.

**PHOENIX**

We calculated a grid of transmission spectra using the PHOENIX atmosphere model[92–94], varying the planet's heat redistribution, atmospheric metallicity, C/O ratio, internal temperature, the presence of aerosols, and the atmospheric chemistry (equilibrium or rainout). Opacities used include the BT2 $H_2O$ line list[95], as well as HITRAN for 129 other major molecular absorbers[96] and Kurucz data for atomic species[97]. The HITRAN line lists available in this version of PHOENIX are often complete only at room temperature, which may be the cause of the apparent shift in the $CO_2$ spectral feature compared to the other grids that primarily use HITEMP and ExoMol lists. This shift is the cause of the difference in $\chi^2$ between PHOENIX and the other model grids. In total, there were 1116 models. Within this grid, the best fit model has 10× solar metallicity, a C/O ratio of 0.3, an internal temperature of 400 K, rainout chemistry, and a cloud deck top at 0.3 mbar. The best fit model has a $\chi^2$/N = 1.203 for N = 344 points.

**PICASO 3.0 & Virga**

We used the open-source radiative-convective equilibrium model PICASO 3.0[98,99], which has its heritage from the Fortran-based EGP mode[100,101], to compute a grid of one-dimensional pressure-temperature models for WASP-39b. The opacity sources included in PICASO 3.0 are listed in Extended Data Table 2. Of the 29 molecular opacity sources included, the line lists of notable molecules used were: $H_2O$[102], $CO_2$[103], $CH_4$[104], CO[105]. The parameters varied in this grid of models include the interior temperature of the planet ($T_{int}$), atmospheric metallicity, C/O ratio, and the dayside-to-nightside heat redistribution factor (see Extended Data Table 2), with correlated-k opacities[106,101]. In total, there were 192 cloud-free models. We include the effect of clouds in two ways. First, we post-processed the pressure-temperature profile using the cloud model Virga[98,107], which follows from previously developed methodologies[39], where we included three condensable species (MnS, $Na_2S$, $MgSiO_3$). Virga requires a vertical mixing parameter, $K_{zz}$ ($cm^2/s$), and a vertically constant sedimentation efficiency parameter, $f_{sed}$. In general, $f_{sed}$ controls the vertical extent of the cloud opacity, with low values ($f_{sed}<1$) creating large vertically extended cloud decks with small particle sizes. In total, there were 3,840 cloudy models. The best fit from our grid with Virga-computed clouds has 3x solar metallicity, solar C/O (0.458), and $f_{sed}$ = 0.6, which results in a $\chi^2/N = 1.084$.

In addition to the grid fit, we also use the PICASO framework to quantify the feature detection significance. In this method, we are able to incorporate clouds on the fly using the fitting routine PyMultiNest[108]. We fit for each of the grid parameters using a nearest neighbour technique and a radius scaling to account for the unknown reference pressure, giving 5 parameters total. When fitting for clouds, we either fit for $K_{zz}$ and $f_{sed}$, within the Virga framework (7 parameters total), or we fit for the cloud top pressure corresponding to a grey cloud deck with infinite opacity (6 parameters total). These results are described in the following section.

**Extended Data Table 2:** The parameter space explored by each RCTE model grid. The best-fit model for each grid is bolded.

| **Grid Name** | **ATMO** | **PHOENIX** | **PICASO+Virga** |
|---|---|---|---|
| Resolution/Sampling | R = 1000 & R = 3000 (Corr-k) | 1 Å sampling | R=60,000 (resampling) |
| Wavelength Range | 0.2 - 30 µm | 0.2-5.35 µm | 0.3-14 µm |
| **Global parameters** | | | |
| Internal temperature (K) | **100**, 200, 300, 400 | 200, **400** | **100**, 200, 300 |
| Heat redistribution | f = 0.25, **0.5**. 0.75, 1.0 (0.5=full, 1 = no redistribution) | f = 0.172, 0.25, **0.351** (0.25 = full redistribution) | f = 0.4, **0.5** (0.5=full redistribution) |
| **Chemistry parameters** | | | |
| Metallicity | 0.1, 1, **3**, 5, 10, 50, 100, 200× solar | 0.1, 1, **10**, 50, 100× solar | 0.1, 0.3, 1, **3**, 10, 30, 50, 100× solar |
| C/O ratio | **0.35**, 0.55, 0.7, 0.75, 1.0, 1.5 | **0.3**, 0.54, 0.7, 1.0 | 0.229, **0.458**, 0.687, 0.916 |
| Elemental Abundance Reference | 109, 110 | 110 | 111 |
| Solar C/O | 0.55 | 0.54 | 0.458 |
| **Aerosol parameters** | | | |
| $f_{sed}$ | N/A | N/A | **0.6**, 1, 3, 6, 10 |
| $K_{zz}$ ($cm^2/s$) | N/A | N/A | 1e5, 1e7, **1e9**, 1e11 |
| Cloud Opacities | Grey (kappa factor - 0.0, 0.5, 1.0, **5.0**) | Grey | MnS, $Na_2S$, $MgSiO_3$, grey |
| $P_{cloud}$ | **1** to 50 mbar (fixed) | None, **0.3**, 1, 3, 10 mbar | Variable (fit on the fly) |
| Rayleigh scattering | **$H_2$ only**, 10× | **$H_2$ only**, 10× | $H_2$ only |
| **Molecular and Atomic Opacity Sources Included** | | | |
| | $CH_4$, CO, $CO_2$, $C_2H_2$, Cs, FeH, HCN, $H_2O$, $H_2S$, K, Li, Na, $NH_3$, $PH_3$, Rb, $SO_2$, TiO, VO | CH, $CH_4$, CN, CO, $CO_2$, COF, $C_2$, $C_2H_2$, $C_2H_4$, $C_2H_6$, CaH, CrH, FeH, HCN, HCl, HF, HI, HDO, $HO_2$, $H_2$, $H_2S$, $H_2O$, $H_2O_2$, $H_3$+, MgH, NH, $NH_3$, NO, $N_2$, $N_2O$, OH, $O_2$, $O_3$, $PH_3$, $SF_6$, SiH, SiO, $SiO_2$, TiH, TiO, VO, atoms up to U | $CH_4$, CO, $CO_2$, $C_2H_2$, $C_2H_4$, $C_2H_6$, CrH, Cs, Fe, FeH, HCN, $H_2$, $H_2O$, $H_2S$, $H_3$+, K, Li, LiCl, LiH, MgH, $NH_3$, $N_2$, Na, OCS, $PH_3$, Rb, SiO, TiO, VO |
| **Statistical Parameters for best-fit (bolded) model** | | | |
| $\chi^2/N$ (N=344) | 1.098 | 1.203 | 1.084 |

**Extended Data Table 3.** Feature detection significance for dominant sources of opacity with two different methods. *B* is the Bayes Factor.

| **Gas** | **Bayesian Gas Removal** | | **Gaussian Residual Fit** |
|---|---|---|---|
| | **ln *B*** | **σ** | **σ** |
| **$H_2O$** | 402.6 | 21.5 | 16.5 |
| **$CO_2$** | 229.0 | 28.5 | 26.9 |
| **$SO_2$** | 9.7 | 4.8 | 3.5 |
| **CO** | -5.0 | 0.3 | 4.5 |

**Feature detection significance**

From the chemical equilibrium results of the single best-fit models, the molecules that could potentially contribute to the spectrum based on their abundances and 3-5 μm opacity sources are $H_2$ & He (via continuum), and CO, $H_2O$, $H_2S$, $CO_2$, and $CH_4$. More minor sources of opacity with VMR abundances of <1 ppm are molecules such as OCS & $NH_3$. For example, removing $H_2S$, $NH_3$, and OCS from the single best fit PICASO 3.0 model increases the chi-squared by less than 0.002. Therefore, we focus on computing the statistical significance of only $H_2O$, $SO_2$, $CO_2$, $CH_4$, and CO.

To quantify the statistical significance, we performed two different tests. First, we used a Gaussian Residual Fitting analysis, as used in other JTEC ERS analyses[23,29,32]. In this method, we subtracted the best-fit model without a specific opacity source from the weighted average spectrum of WASP-39b, isolating the supposed spectral feature. We then fit a 3- or 4-parameter Gaussian curve to the residual data using a nested sampling algorithm to calculate the Bayesian evidence[112]. For $H_2O$ and CO, the extra transit depth offset parameter for the Gaussian fit was necessary to account for local mismatch of the fit to the continuum, while only a mean, standard deviation, and scale parameter were required for a residual fit to the other molecules. We then compared this to the Bayesian evidence of a flat line to find the Bayes factor between a model that fits the spectral feature versus a model that excludes the spectral feature. These fits are shown in Extended Data Figure 6.

While the Gaussian Residual Fitting method is useful for quantifying the presence of potentially unknown spectral features, it cannot robustly determine the source of any given opacity. We therefore used the Bayesian fitting routine from PyMultiNest within the PICASO 3.0 framework to refit the grid parameters, while excluding the opacity contribution from the species in question. Then, we compared the significance of the molecule via a Bayes factor analysis[113]. Those values are shown in Extended Table 3.

We find significant evidence (>3σ) for $H_2O$, $CO_2$, and $SO_2$. In general, the two methods only agree well for molecules whose contribution has a Gaussian shape (i.e. $SO_2$, and $CO_2$). For example, for $CO_2$ we find decisive 28.5σ and 26.9σ detections for the Bayes factor and Gaussian analysis, respectively. Similarly, for $H_2O$, we find 21.5σ and 16.5σ detections, respectively. The evidence for $SO_2$ is less substantial, but both methods give significant detections of 4.8σ and 3.5σ, respectively. While the Gaussian fitting method found a broad 1 μm-wide residual in the region of CO (i.e., >4.5μm), its shape was unlike that seen with the PRISM data[32]. CO remained undetected with the Bayesian fitting analysis and therefore we are unable to robustly confirm evidence of CO. Similarly, no evidence for $CH_4$ was found[23]. Gaussian residual fitting in the region of $CH_4$ absorption only found a very broad inverse Gaussian and so is not included in ED Table 3.

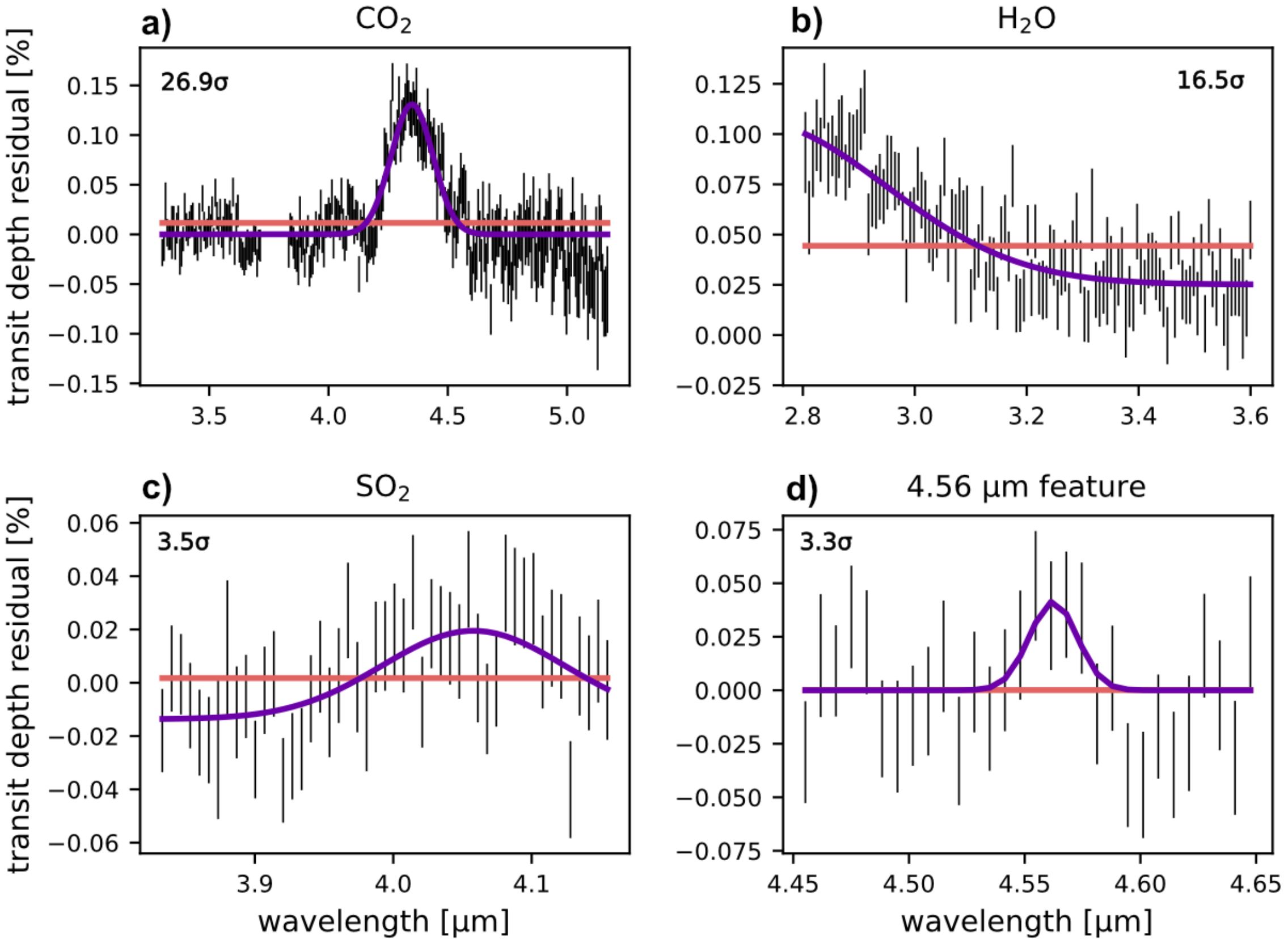

**Extended Data Figure 6: Gaussian versus flat line fits to the residual transmission spectrum for $CO_2$, $H_2O$, $SO_2$, and the 4.56 μm feature** after all other absorption from the best-fit model is subtracted from the data. Each of the Gaussian fits has a higher Bayesian evidence than the flat line fits, indicating a detection, though to varying degrees of significance.

**$SO_2$ Absorption**

We performed an injection test with the PICASO best-fit model within the PyMultiNest fitting framework to determine the abundance of $SO_2$ required to match the observations. We add $SO_2$ opacity using the ExoMol linelist[114], without rerunning the RCTE model to self-consistently compute a new climate profile. Fitting for the cloud deck dynamically, without $SO_2$, produces a single best estimate of 10x solar metallicity, sub-solar C/O (0.229), resulting in a marginally worse $\chi^2/N = 1.11$. With $SO_2$, the single best fit tends back to 3x solar metallicity, solar C/O. This suggests that cloud treatment and the exclusion of spectrally active molecules have an effect on the resultant physical interpretation of bulk atmospheric parameters. Ultimately, if we fit for $SO_2$ within our PyMultiNest framework with the Virga cloud treatment we obtain 3x solar metallicity, solar C/O, log $SO_2$= –5.6 +/- 0.1 (SO2=2.5 ± 0.65 ppm) and $\chi^2/N = 1.02$, which is our single best fit model (shown in Main Text Figure 4). For context, an atmospheric metallicity of 3–10x solar would provide a thermochemical equilibrium abundance of 72-240 ppm $H_2S$, the presumed source for photochemically produced $SO_2$[37].

To confirm the plausibility of $SO_2$ absorption to explain the 4.1 μm spectral feature, we also computed models with prescribed, vertically uniform $SO_2$ VMRs of 0, 1, 5, and 10 ppm using the structure from the best-fit PHOENIX model (10x solar metallicity, C/O = 0.3). We calculated ad-hoc spectra using the gCMCRT radiative transfer code[115] with the ExoMol $SO_2$ linelist[114] (see Extended Data Figure 7). Linearly interpolating the models with respect to the $SO_2$ abundance and performing a Levenberg-Marquardt regression gave a best-fit value of 4.6 +/- 0.67 ppm. Inserting this abundance of $SO_2$ into the best-fit PHOENIX model improves the $\chi^2/N$ from 1.2 to 1.08.

Future atmospheric retrievals can provide a more statistically robust measurement for the $SO_2$ abundance and add additional information from the similar absorption seen in the PRISM transmission spectrum[29,32].

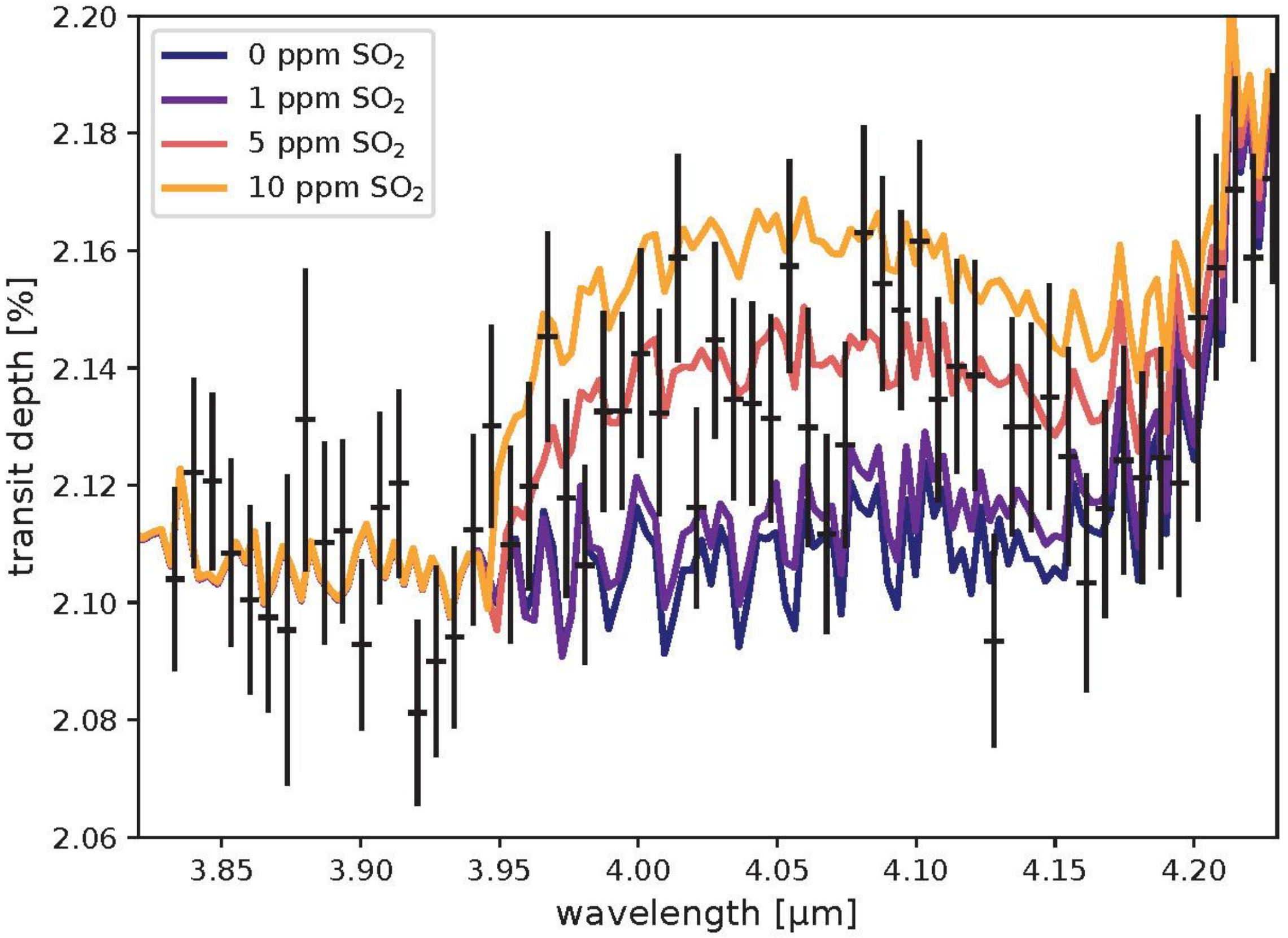

**Extended Data Figure 7: Model transmission spectra of WASP-39b with PHOENIX and gCMCRT with varying abundances of $SO_2$** compared to the observed spectral feature at 4.1 µm in the G395H data. At wavelengths short of 3.95 µm, which is outside of the $SO_2$ band, all models overlap, further suggesting that the data can be explained by the presence of $SO_2$ in the atmosphere. By interpolating these 10x solar metallicity models, we find a best fit $SO_2$ abundance of 4.6 +/- 0.67 ppm. With the best-fit PICASO 3.0 at 3× solar metallicity, we find an $SO_2$ abundance of 2.5 +/- 0.65 ppm.

**4.56 µm Feature**

A 0.08 µm wide bump in transit depth centred at 4.56 µm is not fit by any of the model grids. This feature, picked up by the resolution of G395H, is not clearly seen in other ERS observations of WASP-39b. Following the same Gaussian residual fitting procedure as described above, we found a feature significance of 3.3σ (see Extended Data Figure 6). To identify possible opacity sources in the atmosphere of WASP-39b that might be the cause of this absorption, we compared the feature with $CH_4$[116], $C_2H_2$[117], $C_2H_4$[118], $C_2H_6$[119], CO[120], $CO_2$[103], $CS_2$[119], CN[121], HCN[122], HCl[119], $H_2S$[123], HF[124], $H_3$+[125], LiCl[121], $NH_3$[126], NO[120], $NO_2$[119], $N_2O$[120], $N_2$[127], NaCl[128], OCS[119], $PH_3$[129], PN[130], PO[131], SH[132], SiS[133], $SiH_4$[134], SiO[135], the X-X state of SO[136], $SO_2$[114], $SO_3$[114], and isotopologues of $H_2O$, $CH_4$, $CO_2$, and CO, but did not find a convincing candidate that showed opacity at the correct wavelength or the correct width. The narrowness of the feature suggests it could be a very distinct Q-branch, where the rotational quantum number in the ground state is the same as the rotational quantum number in the excited state. However, of the molecules we explored, there were no candidates with a distinct Q-branch at this wavelength whose P- and R- branches did not obstruct the neighbouring $CO_2$ and continuum-like $CO+H_2O$ opacity.

We also note that many of these species lack high-temperature linelist data, making it difficult to definitively rule out such species. For example, OCS, SO, and $CS_2$ are available in HITRAN 2020[119] but not in ExoMol[137]. Additionally, if photochemistry is significant for WASP-39b as indicated by the presence of $SO_2$, then there may be many species out-of-equilibrium that may contribute to the transit spectrum, some of which do not currently have high-temperature opacity data (e.g., OCS, $NH_2$, HSO). Future observations over this wavelength region of this and other planets may confirm or refute the presence of this unknown absorber.

## Data Availability

The data used in this paper are associated with JWST program ERS 1366 (observation #4) and are available from the Mikulski Archive for Space Telescopes (https://mast.stsci.edu). Science data processing version (SDP_VER) 2022_2a generated the uncalibrated data that we downloaded from MAST. We used JWST Calibration Pipeline software version (CAL_VER) 1.5.3 with modifications described in the text. We used calibration reference data from context (CRDS_CTX) 0916, except as noted in the text. All the data and models presented in this publication can be found at 10.5281/zenodo.7185300.

## Code Availability

The codes used in this publication to extract, reduce and analyze the data are as follows; STScI JWST Calibration pipeline[45] (https://github.com/spacetelescope/jwst), Eureka![53] (https://eurekadocs.readthedocs.io/en/latest/), ExoTiC-JEDI[47] (https://github.com/Exo-TiC/ExoTiC-JEDI), juliet[71] (https://juliet.readthedocs.io/en/latest/), Tiberius[15,49,50], transitspectroscopy[51] (https://github.com/nespinoza/transitspectroscopy). In addition, these made use of batman[65] (http://lkreidberg.github.io/batman/docs/html/index.html), celerite[86] (https://celerite.readthedocs.io/en/stable/), chromatic (https://zkbt.github.io/chromatic/), Dynesty[72] (https://dynesty.readthedocs.io/en/stable/index.html), emcee[69] (https://emcee.readthedocs.io/en/stable/), exoplanet[83] (https://docs.exoplanet.codes/en/latest/), ExoTEP[75–77], ExoTHETyS[79] (https://github.com/ucl-exoplanets/ExoTETHyS), ExoTiC-ISM[57] (https://github.com/Exo-TiC/ExoTiC-ISM), ExoTiC-LD[58] (https://exotic-ld.readthedocs.io/en/latest/), george[68] (https://george.readthedocs.io/en/latest/) JAX[82] (https://jax.readthedocs.io/en/latest/), LMFIT[70] (https://lmfit.github.io/lmfit-py/), Pylightcurve[78] (https://github.com/ucl-exoplanets/pylightcurve), Pymc3[138] (https://docs.pymc.io/en/v3/index.html) and Starry[84] (https://starry.readthedocs.io/en/latest/), each of which use the standard python libraries astropy[139,140], matplotlib[141], numpy[142], pandas[143], scipy[64] and xarray[144]. The atmospheric models used to fit the data can be found at ATMO[Tremblin2015,Drummond2016,Goyal2018,Goyal2020][88–91], PHOENIX[92–94], PICASO[98,99] (https://natashabatalha.github.io/picaso/), Virga[98,107] (https://natashabatalha.github.io/virga/), and gCMCRT[115] (https://github.com/ELeeAstro/gCMCRT).

**Methods References**

**Acknowledgements**
This work is based on observations made with the NASA/ESA/CSA James Webb Space Telescope. The data were obtained from the Mikulski Archive for Space Telescopes at the Space Telescope Science Institute, which is operated by the Association of Universities for Research in Astronomy, Inc., under NASA contract NAS 5-03127 for JWST. These observations are associated with program JWST-ERS-01366. Support for program JWST-ERS-01366 was provided by NASA through a grant from the Space Telescope Science Institute, which is operated by the Association of Universities for Research in Astronomy, Inc., under NASA contract NAS 5-03127. LA acknowledges funding from STFC grant ST/W507337/1 and from the University of Bristol School of Physics PhD Scholarship Fund.

**Author contributions**
All authors played a significant role in one or more of the following: development of the original proposal, management of the project, definition of the target list and observation plan, analysis of the data, theoretical modelling, and preparation of this manuscript.
Some specific contributions are listed as follows. NMB, JLB, and KBS provided overall program leadership and management. LA and HRW led the efforts for this manuscript. DS, EK, HRW, IC, JLB, KBS, LK, MLM, MRL, NMB, VP, and ZBT made significant contributions to the design of the program. KBS generated the observing plan with input from the team. EvSc, NE, and TGB provided instrument expertise. BB, EK, HRW, IC, JLB, LK, MLM, MRL, NMB, and ZBT led or co-led working groups and/or contributed to significant strategic planning efforts like the design and implementation of the pre-launch Data Challenges. AC, DS, EvSc, NE, NG, TGB, VP generated simulated data for pre-launch testing of methods. LA, HRW, MKA, NEB, and JDL contributed significantly to the writing of this manuscript, along with contributions in the Methods from JA, SB, MD, NE, LF, JG, DG, JI, TME, PAR, NW. LA, HRW, MKA, JA, SB, MD, NE, LF, DG, JI, TME, PAR and NW contributed to the development of data analysis pipelines and/or provided the data analysis products used in this analysis i.e., reduced the data, modeled the light curves, and/or produced the planetary spectrum, with additional contributions from JB, TD and LRR. JDL, NEB, JG, EL, and RH generated theoretical model grids for comparison with data. HRW, JDL, and NEB generated figures for this manuscript. MLM, KC, NG, LK, ML, JM and EvSc provided significant feedback to the manuscript coordinating comments from all other authors.

**Competing interests** The authors declare no competing interests.

**Author affiliations**
[1]School of Physics, HH Wills Physics Laboratory, University of Bristol, Bristol, UK
[2]Earth and Planets Laboratory, Carnegie Institution for Science, Washington, DC, USA

[3]NASA Ames Research Center, Moffett Field, CA, USA
[4]Department of Physics, Utah Valley University, Orem, UT, USA
[5]Center for Astrophysics | Harvard & Smithsonian, Cambridge, MA, USA
[6]Anton Pannekoek Institute for Astronomy, University of Amsterdam, Amsterdam, The Netherlands
[7]Department of Physics & Astronomy, University of Kansas, Lawrence, KS, USA
[8]Astrophysics Section, Jet Propulsion Laboratory, California Institute of Technology, Pasadena, CA, USA
[9]Department of Astrophysical Sciences, Princeton University, Princeton, NJ, USA
[10]LSSTC Catalyst Fellow
[11]Space Telescope Science Institute, Baltimore, MD, USA
[12]Department of Physics & Astronomy, Johns Hopkins University, Baltimore, MD, USA
[13]Department of Astronomy and Carl Sagan Institute, Cornell University, Ithaca, NY, USA
[14]School of Earth and Planetary Sciences (SEPS), National Institute of Science Education and Research (NISER), HBNI, Odisha, India
[15]Division of Geological and Planetary Sciences, California Institute of Technology, Pasadena, CA, USA
[16]Center for Space and Habitability, University of Bern, Bern, Switzerland
[17]Max Planck Institute for Astronomy, Heidelberg, Germany
[18]Department of Physics and Institute for Research on Exoplanets, Université de Montréal, Montreal, QC, Canada
[19]Department of Astronomy & Astrophysics, University of California, Santa Cruz, Santa Cruz, CA, USA
[20]Department of Astronomy & Astrophysics, University of Chicago, Chicago, IL, USA
[21]Department of Astrophysical and Planetary Sciences, University of Colorado, Boulder, CO, USA
[22]European Space Agency, Space Telescope Science Institute, Baltimore, MD, USA
[23]Department of Physics and Astronomy, University College London, United Kingdom
[24]NASA Goddard Space Flight Center, Greenbelt, MD, USA
[25]Center for Computational Astrophysics, Flatiron Institute, New York, New York, USA
[26]School of Physics, Trinity College Dublin, Dublin, Ireland
[27]School of Earth and Space Exploration, Arizona State University, Tempe, AZ, USA
[28]University Observatory Munich, Ludwig Maximilian University, Munich, Germany
[29]Exzellenzcluster Origins, Garching, Germany
[30]Lunar and Planetary Laboratory, University of Arizona, Tucson, AZ, USA.
[31]Instituto de Astrofísica de Canarias (IAC), Tenerife, Spain
[32]Departamento de Astrofísica, Universidad de La Laguna (ULL), Tenerife, Spain
[33]INAF- Palermo Astronomical Observatory, Piazza del Parlamento, Palermo, Italy
[34]Space Science Institute, Boulder, CO, USA
[35]Steward Observatory, University of Arizona, Tucson, AZ, USA
[36]Department of Earth and Planetary Sciences, Johns Hopkins University, Baltimore, MD, USA
[37]Johns Hopkins APL, Laurel, MD, USA
[38]Atmospheric, Oceanic and Planetary Physics, Department of Physics, University of Oxford, Oxford, UK

[39]Indian Institute of Technology, Indore, India
[40]Centre for Exoplanets and Habitability, University of Warwick, Coventry, UK
[41]Department of Physics, University of Warwick, Coventry, UK
[42]NSF Graduate Research Fellow
[43]School of Physical Sciences, The Open University, Milton Keynes, UK
[44]BAER Institute, NASA Ames Research Center, Moffet Field, CA, USA
[45]Department of Physics, New York University Abu Dhabi, Abu Dhabi, UAE
[46]Center for Astro, Particle and Planetary Physics (CAP3), New York University Abu Dhabi, Abu Dhabi, UAE
[47]School of Physics and Astronomy, University of Leicester, Leicester
[48]Centre for Exoplanet Science, University of St Andrews, St Andrews, UK
[49]Leiden Observatory, University of Leiden, Leiden, The Netherlands
[50]INAF – Osservatorio Astrofisico di Torino, Pino Torinese, Italy
[51]Space Research Institute, Austrian Academy of Sciences, Graz, Austria
[52]Institute of Astronomy, Department of Physics and Astronomy, KU Leuven, Leuven, Belgium
[53]Planetary Sciences Group, Department of Physics and Florida Space Institute, University of Central Florida, Orlando, Florida, USA
[54]Universitäts-Sternwarte, Ludwig-Maximilians-Universität München, München, Germany
[55]Institute for Astrophysics, University of Vienna, Vienna, Austria
[56]Department of Astronomy, University of Maryland, College Park, MD, USA
[57]Department of Physics, Imperial College London, London, UK
[58]Imperial College Research Fellow
[59]California Institute of Technology, Pasadena, CA, USA
[60]Laboratoire d'Astrophysique de Bordeaux, Université de Bordeaux, Pessac, France
[61]Département d'Astronomie, Université de Genève, Sauverny, Switzerland
[62]Department of Astronomy, University of Michigan, Ann Arbor, MI, USA
[63]NHFP Sagan Fellow
[64]Department of Physics, University of Rome "Tor Vergata", Rome, Italy
[65]INAF - Turin Astrophysical Observatory, Pino Torinese, Italy
[66]Department of Physics and Astronomy, Faculty of Environment Science and Economy, University of Exeter, EX4 4QL, UK
[67]SRON Netherlands Institute for Space Research, Leiden, the Netherlands
[68]Université Côte d'Azur, Observatoire de la Côte d'Azur, CNRS, Laboratoire Lagrange, France
[69]Department of Earth, Atmospheric and Planetary Sciences, Massachusetts Institute of Technology, Cambridge, MA, USA
[70]Kavli Institute for Astrophysics and Space Research, Massachusetts Institute of Technology, Cambridge, MA, USA
[71]51 Pegasi b Fellow
[72]Astronomy Department and Van Vleck Observatory, Wesleyan University, Middletown, CT, USA
[73] Institute of Astronomy, University of Cambridge, Cambridge, UK
[74]Maison de la Simulation, CEA, CNRS, Univ. Paris-Sud, UVSQ, Université Paris-Saclay, Gif-sur-Yvette, France

[75]Planetary Science Institute, Tucson, AZ, USA
[76]Université de Paris Cité and Univ Paris Est Creteil, CNRS, LISA, Paris, France
[77]Department of Earth and Planetary Sciences, University of California Santa Cruz, Santa Cruz, California, USA